\documentclass[a4paper,fleqn,usenatbib]{mnras}

\usepackage[T1]{fontenc}
\usepackage{ae,aecompl}

\usepackage{graphicx}	% Including figure files
\usepackage{amsmath}	% Advanced maths commands
\usepackage{amssymb}	% Extra maths symbols
\usepackage{array}
\usepackage{xcolor}

\title[MaNGA {\sc firefly} VAC]{The MaNGA {\sc firefly} Value-Added-Catalogue: resolved stellar populations of 10,010 nearby galaxies}

\author[J. Neumann et al.]{
Justus Neumann,$^{1}$\thanks{E-mail: jusneuma.astro@gmail.com}
Daniel Thomas,$^{1,2}$
Claudia Maraston,$^{1}$
Lewis Hill,$^{1}$\newauthor
Lorenza Nanni,$^{1}$
Oliver Wenman,$^{1}$
Jianhui Lian,$^{3}$
Johan Comparat,$^{4}$\newauthor
Violeta Gonzalez-Perez,$^{5,6}$
Kyle B. Westfall,$^{7}$
Renbin Yan,$^{8}$
Yanping Chen,$^{9}$\newauthor
Guy S. Stringfellow,$^{10}$
Matthew A. Bershady,$^{11,12,13}$
Joel R. Brownstein,$^{14}$\newauthor
Niv Drory,$^{15}$
Donald P. Schneider$^{16,17}$
\\
$^{1}$Institute of Cosmology and Gravitation, University of Portsmouth, Burnaby Road, Portsmouth, PO1 3FX, UK\\
$^{2}$School of Mathematics and Physics, University of Portsmouth, Lion Gate Building, Portsmouth, PO1 3HF, UK\\
$^{3}$Max-Planck-Institut f\"ur Astronomie, K\"onigstuhl 17, D-69117 Heidelberg, Germany\\
$^{4}$Max-Planck-Institut f\"ur Extraterrestrische Physik (MPE), Giessenbachstrasse 1, 85748, Garching bei M\"unchen, Germany\\
$^{5}$Departamento de F\'isica Te\'orica, M\'odulo 15, Facultad de Ciencias, Universidad Aut\'onoma de Madrid, E-28049 Madrid, Spain\\
$^{6}$Centro de Investigaci\'on Avanzada en F\'isica Fundamental, Facultad de Ciencias, Universidad Aut\'onoma de Madrid,\\
E-28049 Madrid, Spain\\
$^{7}$University of California Observatories, University of California, Santa Cruz, 1156 High Street, Santa Cruz, CA 95064, USA\\
$^{8}$Department of Physics, The Chinese University of Hong Kong, Shatin, N.T., Hong Kong SAR, China\\
$^{9}$New York University Abu Dhabi, Abu Dhabi, PO Box 129188, United Arab Emirates\\
$^{10}$Center for Astrophysics and Space Astronomy, Department of Astrophysical and Planetary Sciences, University of Colorado,\\
389 UCB, Boulder,80309-0389, USA\\
$^{11}$South African Astronomical Observatory, PO Box 9, Observa- tory 7935, Cape Town, South Africa\\
$^{12}$Department of Astronomy, University of Cape Town, Private Bag X3, Rondebosch 7701, South Africa\\
$^{13}$Department of Astronomy, University of Wisconsin-Madison, 475 N. Charter Street, Madison, WI 53706, USA\\
$^{14}$Department of Physics \& Astronomy, University of Utah, Salt Lake City, UT 84112, USA\\
$^{15}$Department of Astronomy, University of Texas at Austin, Austin, TX 78712, USA\\
$^{16}$Department of Astronomy and Astrophysics, The Pennsylvania State University, University Park, PA 16802, USA\\
$^{17}$Institute for Gravitation and the Cosmos, Pennsylvania State University, University Park, PA 16802, USA
}

\date{Accepted XXX. Received YYY; in original form ZZZ}

\pubyear{2015}

\begin{document}
\label{firstpage}
\pagerange{\pageref{firstpage}--\pageref{lastpage}}
\maketitle

\begin{abstract}

We present the \texttt{MaNGA {\sc firefly} Value-Added-Catalogue (VAC)} -- a catalogue of $\sim 3.7$ million spatially resolved stellar population properties across 10,010 nearby galaxies from the final data release of the MaNGA survey. The full spectral fitting code {\sc firefly} is employed to derive parameters such as stellar ages, metallicities, stellar and remnant masses, star formation histories, star formation rates and dust attenuation. In addition to Voronoi-binned measurements, our \texttt{VAC} also provides global properties, such as central values and radial gradients. Two variants of the \texttt{VAC} are available: presenting the results from fits using the \texttt{M11-MILES} and the novel \texttt{MaStar} stellar population models. \texttt{MaStar} allows to constrain the fit over the whole MaNGA wavelength range, extends the age-metallicity parameter space, and uses empirical spectra from the same instrument as MaNGA. The fits employing \texttt{MaStar} models find on average slightly younger ages, higher mass-weighted metallicities and smaller colour excesses. These differences are reduced when matching wavelength range and converging template grids. We further report that {\sc firefly} stellar masses are systematically lower by $\sim 0.3$\,dex than masses from the MaNGA \texttt{PCA} and \texttt{Pipe3D} VACs, but match masses from the \texttt{NSA} best with only $\sim 0.1$\,dex difference. Finally, we show that {\sc firefly} stellar ages correlate with spectral index age indicators H$\delta_A$ and D$_n(4000)$, though with a clear additional metallicity dependence.

\end{abstract}


%%%%%%%%%%%%%%%%%%%%%%%%%%%%%%%%%%%%%%%%%%%%%%%%%%

%%%%%%%%%%%%%%%%% BODY OF PAPER %%%%%%%%%%%%%%%%%%

\section{Introduction}

Stellar populations are one of the most important fossil records of galaxy evolution. They provide information about mass assembly and chemical enrichment histories including not only the integrated star formation and internal recycling processes but also pristine gas inflows, metal-loss in outflows, or mass and metal accretion through minor mergers. Observations and detailed studies of stellar populations are indispensable to constrain evolutionary processes included in models and cosmological simulations.

Single stars in galaxies outside the Local Group cannot (yet) be resolved and, thus, analyses depend on observations of integrated light. It has become a common approach to approximate the sum of all stellar light in one resolution element by a combination of single-age, single-metallicity stellar populations. Models of these simple stellar populations (SSPs) are synthesised from stellar libraries together with stellar evolution theory in the form of an initial mass function (IMF), stellar tracks and/or isochrones \citep{Bruzual1993,Worthey1994,Vazdekis1996,Fioc1997,Maraston1998,Leitherer1999,
Bruzual2003,Maraston2005,Conroy2009,Maraston2011,Maraston2020}. The adopted library can be either fully theoretical \citep{Maraston1998}, fully empirical \citep{Vazdekis2010,Vazdekis2016,Maraston2020} or a combination of both \citep{Leitherer1999,Maraston2005,Walcher2009,Maraston2011,Conroy2018}. Complex stellar populations are then obtained by combining SSPs with arbitrary star formation histories.

In order to derive stellar population properties of data, stellar population models need to be matched to observed spectra. Over the last decades a variety of codes have been developed to perform full spectral fitting of observed spectra with varying focus on emission lines, stellar kinematics, stellar population or all-in-one approaches: e.g. {\sc pPXF} \citep{Cappellari2004,Cappellari2017}, {\sc starlight} \citep{CidFernandes2005}, {\sc steckmap} \citep{Ocvirk2006a,Ocvirk2006b}, {\sc gandalf} \citep{Sarzi2006,Falcon-Barroso2006}, {\sc vespa} \citep{Tojeiro2007}, {\sc ULySS} \citep{Koleva2009} {\sc PyParadise} \citep{Walcher2015,Husemann2016b}, {\sc beagle} \citep{Chevallard2016}, {\sc fado} \citep{Gomes2017}, {\sc firefly} \citep{Wilkinson2017}, {\sc bagpipes} \citep{Carnall2018} and {\sc prospoector} \citep{Johnson2021}.

With the advent of large-scale integral field unit (IFU) surveys, such as 
SAURON (\citealp{deZeeuw2002}, 48 galaxies), $\rm ATLAS^{3D}$ (\citealp{Cappellari2011}, 260 galaxies), CALIFA (\citealp{Sanchez2012}, 667 galaxies), SAMI (\citealp{Croom2012,Bryant2015}, 3068 galaxies), $\rm KMOS^{3D}$ (\citealp{Wisnioski2015}, 739 galaxies) and MaNGA (\citealp{Bundy2015}, 10010 galaxies), a massive amount of data has been produced for hundreds to thousands of galaxies each including thousands of spectra. Consequently, spectral fitting pipelines have been created to automatize and facilitate the process of creating 2D maps of physical galaxy parameters from observations in form of 3D datacubes (e.g. {\sc Pipe3D}, \citealp{Sanchez2016}, {\sc pyPipe3D}, \citealp{Lacerda2022}, {\sc lzifu}, \citealp{Ho2016}, {\sc gist}, \citealp{Bittner2019}, and the MaNGA data analysis pipeline, {\sc dap}, \citealp{Westfall2019}).

The Mapping Nearby Galaxies at Apache Point Observatory survey \citep[MaNGA;][]{Bundy2015}, a Sloan Digital Sky Survey-IV project \citep[SDSS-IV][]{Blanton2017}, is the largest IFU galaxy survey to date. The final MaNGA data release \citep[SDSS-DR17;][]{Abdurrouf2021} contains observations of 10,010 galaxies. MaNGA data are a massive ressource for studies of galaxy evolution in the nearby universe.

In this paper, we present the \texttt{MaNGA {\sc firefly} Value-Added-Catalogue} (VAC): a catalogue of spatially resolved stellar population parameters for all MaNGA galaxies. The pipeline to create the \texttt{VAC} is built upon the {\sc dap} and uses its output for further processing. The data products contained in the \texttt{VAC} are complementary to the high-level output of the {\sc dap}, which includes stellar kinematics, absorption line indices and emission line measurements, detailed in Sect. \ref{sect:input_dap}. Earlier version of the \texttt{MaNGA {\sc firefly} VAC} have been described and used in \citet{Goddard2017} and \citet{Parikh2018}. Here, we present the first dedicated \texttt{VAC} paper based on the final MaNGA data release and providing a full description and detailed analysis of its content. This version of the \texttt{VAC} has already been used for scientific analysis in \citet{Neumann2021} to investigate the drivers of stellar metallicity in galaxies, showing the scientific potential of this large stellar population database. 

We describe the input and the pre-processing of MaNGA data in Sect. \ref{sect:input}, the workflow to produce the \texttt{VAC} and details about the {\sc firefly} routine are presented in Sect. \ref{sect:workflow}. Data products are shown, tested, discussed in Sect. \ref{sect:output}, and compared to other catalogues in \ref{sect:literature}. We conclude in Sect. \ref{sect:summary}.

Throughout the paper we adopt a flat $\Lambda$CDM cosmology with a Hubble constant of $H_0 = 67.8\mathrm{\,km\,s^{-1}\,Mpc^{-1}}$ and $\Omega_\mathrm{m} = 0.308$ \citep{Planck2016}.

\begin{figure*}
		\centering
		\includegraphics[width=18cm]{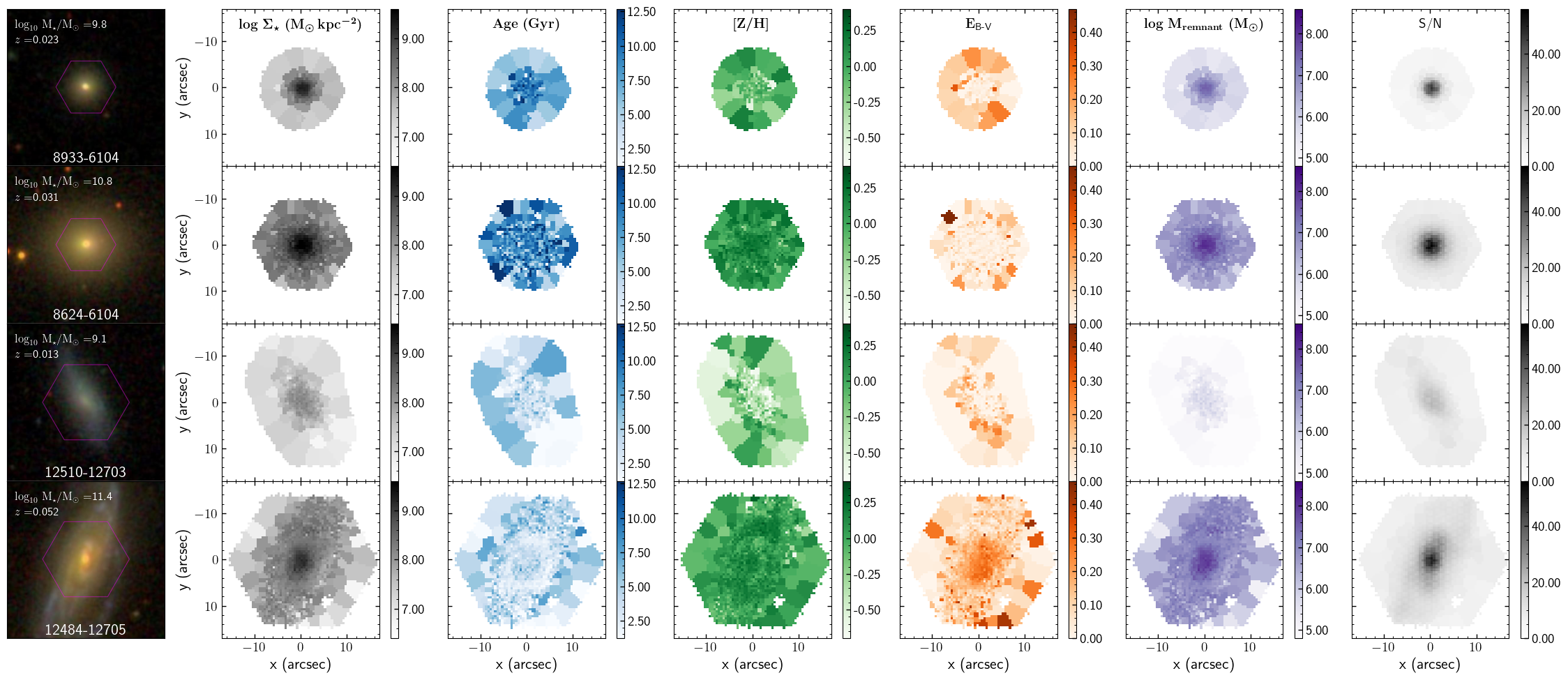}
   		\caption{Example 2D maps from the \texttt{FF-Mi} variant of the \texttt{MaNGA {\sc firefly} VAC} showing different stellar population parameters, dust attenuation and signal-to-noise.}
   		\label{fig:maps}
\end{figure*}

The \texttt{MaNGA {\sc firefly} VAC} can be downloaded as {\sc fits} file from the SDSS website \url{https://www.sdss.org/dr17/manga/manga-data/manga-firefly-value-added-catalog} or from the ICG Portsmouth's website \url{http://www.icg.port.ac.uk/manga-firefly-vac}. It is also available as {\sc cas} table on the SDSS skyserver \url{http://skyserver.sdss.org/dr17} and integrated in {\sc marvin} \url{https://dr17.sdss.org/marvin} \citep{Cherinka2019}. Examples of 2D maps of fitted stellar population parameter available in the \texttt{VAC} are presented in Fig. \ref{fig:maps}.

\section{Input}
\label{sect:input}

The \texttt{MaNGA {\sc firefly} VAC} is a catalogue of spatially resolved stellar population parameters for MaNGA galaxies and, as such, is based entirely on observation of the MaNGA galaxy survey. We employ the full spectral fitting code {\sc firefly} \citep{Wilkinson2017} to fit linear combinations of stellar population models to the observed, reduced and pre-analysed MaNGA spectra. Hence, the two sources of input for the \texttt{MaNGA {\sc firefly} VAC} are the MaNGA data, on the one hand, and the stellar population model libraries, on the other hand.

	\subsection{MaNGA data}
	
		The sample of galaxies for this catalogue is drawn from the 17th and final data release\footnote{\url{https://www.sdss.org/dr17}} \citep[DR17;][]{Abdurrouf2021} of the Mapping Nearby Galaxies at Apache Point Observatory survey \citep[MaNGA;][]{Bundy2015}, a Sloan Digital Sky Survey-IV project \citep[SDSS-IV][]{Blanton2017}. MaNGA uses integral field units (IFUs) that consist of 17 fibre bundles of hexagonal shape ranging from 19 to 127 fibres per bundle in order to ensure a uniform spatial coverage per galaxy \citep{Drory2015}. The IFU fibre bundles are plugged onto observation plates and feed into the BOSS Spectrographs \citep{Smee2013} mounted at the Sloan Foundation 2.5-meter Telescope \citep{Gunn2006}.
		
	Each MaNGA plate is observed following a three-point dithering pattern	to fill the inter-fibre regions and obtain uniform spatial coverage. The observations are repeated for a total observation time of $\sim 2$-$3\,\mathrm{h}$ until the signal-to-noise (S/N) per pixel goal\footnote{The goal is to have a continuum S/N of 33 per pixel in the r-band stacked between 1\,$R_\mathrm{e}$ and 1.5\,$R_\mathrm{e}$.} is reached \citep{Yan2016b}. Each of the two spectrographs has a red and blue camera with overlapping wavelength range and a total coverage of $3622\,$\AA-$10354\,$\AA$\,$ at a median spectral resolution of $\rm \sigma = 72\,km\,s^{-1}$. The median spatial resolution is $2.54\,\arcsec$ full width at half maximum (FWHM), which corresponds to $1.8\,\mathrm{kpc}$ at the median redshift of $z\sim0.037$. \citep{Law2016}.
	
	The main MaNGA sample consists of a Primary sample ($\sim 50\%$), a Colour-Enhanced supplement ($\sim 17\%$) and a higher redshift Secondary sample ($\sim 33\%$). The main selection criteria is a flat stellar mass distribution $M_\star>10^9\,M_\odot$ in logarithmic space with a uniform spatial coverage out to $1.5\,R_\mathrm{e}$ for the Primary sample and $2.5\,R_\mathrm{e}$ for the Secondary sample, where $R_\mathrm{e}$ is the effective radius of a given galaxy. The Colour-Enhanced supplement fills poorly sampled regions in the colour-magnitude diagram \citep{Law2015,Wake2017} and it covers out to $1.5\,R_\mathrm{e}$. In addition to the main sample, the full catalogue also includes observations from ancillary programs\footnote{\url{https://www.sdss.org/dr17/manga/manga-target-selection/ancillary-targets/}}, which make about $\sim 5\%$ of the total sample and are processed by {\sc firefly}, as well. These include single galaxy observations as well as mosaicing of larger regions from a variety of science programmes.
	
	Observations have been concluded in August 2020 and the final sample has been released in \citet{Abdurrouf2021}. The total number of datacubes is 11273 including 10145 galaxy observations of high quality corresponding to 10010 unique galaxies. The \texttt{MaNGA {\sc firefly} VAC} provides the stellar populations for 10735 of these datacubes.\footnote{The \texttt{MaNGA {\sc firefly} VAC} processes all datacubes that successfully run through the {\sc dap}. From the total number of 11273 datacubes, 538 failed during the {\sc dap} run due to various reasons.}	
	
	All raw data are processed by two survey pipelines: the MaNGA data reduction pipeline \citep[\texttt{DRP;}][]{Law2016} and the MaNGA data analysis pipeline \citep[\texttt{DAP;}][]{Westfall2019} to provide high-level science-ready data products, which are essential inputs for the {\sc firefly} \texttt{VAC} project.
	
	\subsection{Data reduction pipeline}
	
	The main task of the \texttt{DRP} is to read the raw data from MaNGA observations and extract, calibrate and combine them into FITS files that are ready for scientific analysis. This is a sophisticated multi-step process that is described in full detail in \citet{Law2016} and updates to DR15 in \citet{Aguado2019} and to DR17 in \citet{Abdurrouf2021}. In addition, the detailed modelling of the line-spread function (LSF) is described in \citet{Law2021}. Here, we provide only a very short summary.
	
	First, the flux of each fibre is extracted, flatfielded and wavelength calibrated using specific calibration frames. The background sky is then determined from a combination of eight dedicated sky fibres and subtracted from each of the science fibre spectra. Afterwards, each spectrum is flux calibrated employing standard star observations from 12 small 7-fibre bundles with the same fibre size and fill-factor as the science fibre bundles \citep{Yan2016a}. The two spectra from the blue and red camera are then combined into a single spectrum and resampled onto a common wavelength array. The output of the first stage is a FITS file containing row-stacked spectra (RSS) of one fully reduced spectrum per fibre. During the second stage, the astrometry of each exposure is registered and the series of 2D RSS are combined and resampled into a 3D datacube with a grid of $0\farcs5$ per pixel. The \texttt{DRP} produces one datacube for each `PLATEIFU'\footnote{Each MaNGA target is identified by its own `MaNGA-ID', but some targets are observed multiple times. `PLATEIFU' is a string with a unique plate-IFU combination for a given observation} designation.

	\subsection{Data analysis pipeline}
	\label{sect:input_dap}
	
	Every datacube that has been produced by the \texttt{DRP} and that has an initial redshift estimate is fed into the \texttt{DAP} for the analysis of higher-level data products. In addition to the \texttt{DRP} datacubes as input, the \texttt{DAP} uses photometric measurements of the ellipticity and position angle of the target from the parent catalogue \citep{Wake2017}; the enhanced NASA Sloan Atlas\footnote{M. Blanton; \url{www.nsatlas.org}}. For each datacube, the \texttt{DAP} produces spatially resolved stellar kinematics, emission line properties and spectral indices. In the following, we provide a short outline of the \texttt{DAP} workflow, full details can be found in \citet{Westfall2019} and for the emission line fitting in \citet{Belfiore2019}.
	
	The pipeline is executed in a series of six main modules that are tasked for: \texttt{DRP} output assessment, spatial binning, stellar kinematics fitting, emission line moments fitting, Gaussian emission line modelling, and stellar index measurement. The assessment stage secures that a spectrum is only analysed if 80\% of the data points are valid and, in addition, it computes the $g$-band weighted $S/N$ per spectrum. This is followed by spatial Voronoi binning \citep{Cappellari2003} of the spectra to a minimum target $S/N \sim 10$ per bin. Afterwards, all fitting modules are executed multiple times using different combinations of binned and unbinned data, the name of each approach is saved in the keyword `DAPTYPE'. For DR17, the \texttt{DAP} provides output products for analysis of single spaxels (`SPX'), of Voronoi-binned spectra (`VOR10'), and of a hybrid binning scheme, where the stellar continuum is fitted on the binned spectra while the emission lines are measured per spaxel (`HYB10'). Similar to the fitting of stellar kinematics, stellar population analysis usually requires higher S/N in the continuum than emission line analysis due to the much stronger signal in the emission lines. The \texttt{MaNGA {\sc firefly} VAC} therefore uses exclusively the `VOR10' \texttt{DAP} output.
	
	The stellar kinematics are fitted employing the penalized pixel-fitting method \citep[\texttt{pPXF};][]{Cappellari2004,Cappellari2017} with a hierarchically clustered selection of stellar templates from the MILES library \citep{Sanchez-Blazquez2006}, providing a higher spectral resolution. For the stellar continuum modelling in the emission line module the \texttt{DAP} uses a different library: with the advantage of full wavelength coverage, in DR17, a subset of the MaStar single stellar population (SSP) library \citep{Maraston2020} is used. The last stage is the measurement of stellar indices on the emission line subtracted spectra. These include absorption line indices and bandhead indices.
	
	The two main output products are a `MAPS' file, which contains 2D maps of high-level data products, and a `LOGCUBE' file, which contains the original and best-fitting model spectra as well as emission line spectra for each \texttt{DRP} datacube (observation/`PLATEIFU'). In addition, there is a single summary `DAPall' catalogue that provides global galaxy properties.
	
	The main source of input to the \texttt{MaNGA {\sc firefly} VAC} are the \texttt{DAP} data products. In particular, the \texttt{MaNGA {\sc firefly} VAC} uses for each datacube per Voronoi bin: coordinates, $g$-band S/N, velocity, velocity dispersion (from MAPS file), observed spectrum, emission line spectrum, bitmask, inverse variance, line spread function (from LOGCUBE file), and redshift (from DAPall file).

	\subsection{Stellar population model libraries}
	\label{sect:ssps}
	
	The current version of the \texttt{MaNGA {\sc firefly} VAC} released together with SDSS DR17 is offered in two structurally identical variants with the only difference being the stellar population model library employed to fit the observed spectra. Both models are based on the same stellar evolution (as in \citealt{Maraston2005}) and IMF assumption and the difference between them is the adopted stellar spectral library.
	
\begin{figure}
	\centering
	\includegraphics[width=\columnwidth]{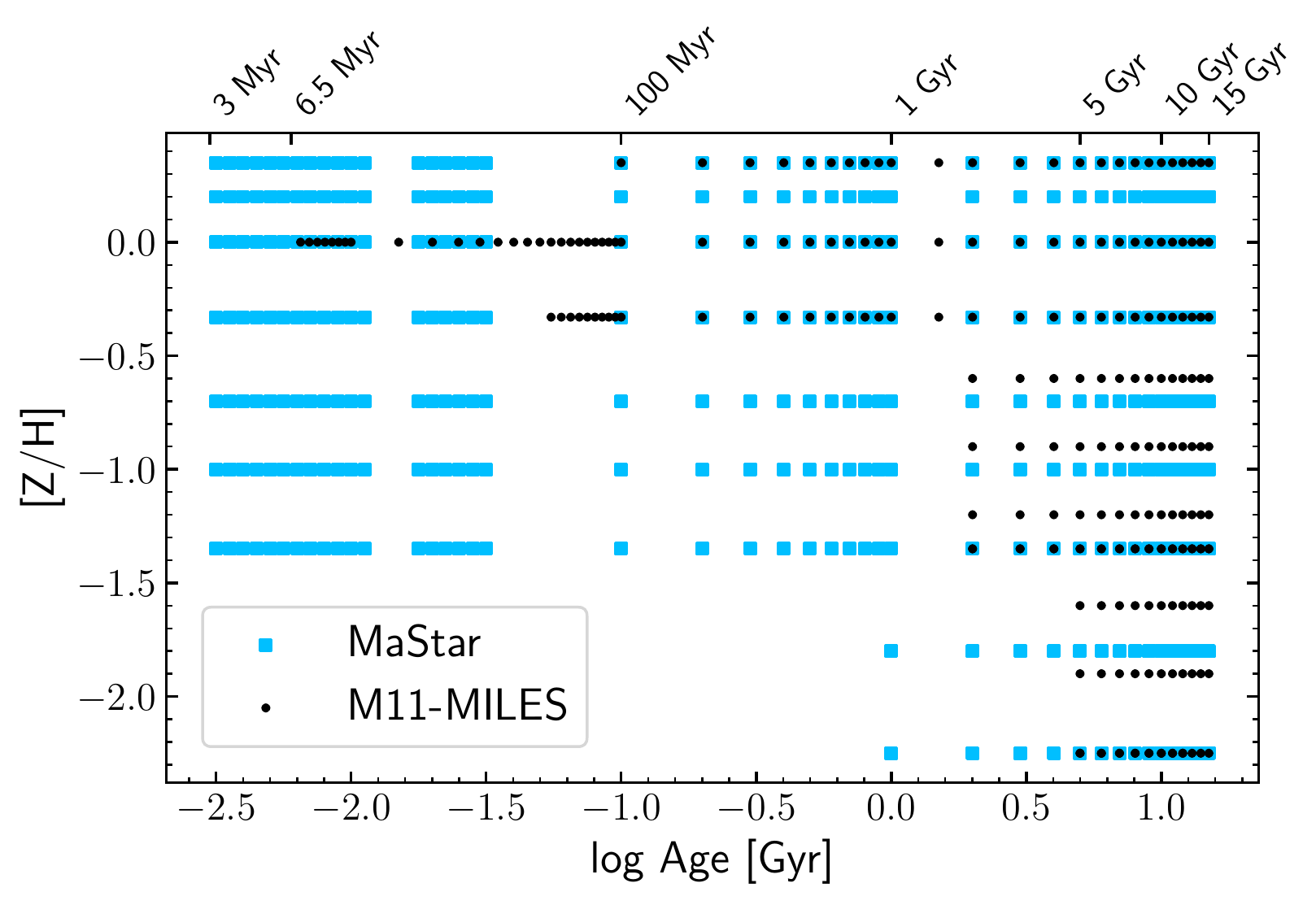}
    \caption{Age-metallicity grid of the two stellar population libraries used in the VAC.}
    \label{fig:grid}
\end{figure}

	The first variant uses the \texttt{M11-MILES} model templates from \citet{Maraston2011} as in the previous versions of the VAC. These single-burst stellar population models are based on the MILES stellar library \citep{Sanchez-Blazquez2006} and inherit its native wavelength coverage, from $3500\,$\AA$\,$ to $7430\,$\AA$\,$, and spectral resolution of $2.54\,$\AA$\,$FWHM \citep{Beifiori2011,FalconBarroso2011}. The population synthesis code adopts isochrones and stellar tracks by \citet{Cassisi1997} for ages older than $30\,\mathrm{Myr}$ and by \citet{Schaller1992} for younger ages. Here, we use the models that are based on a \citet{Kroupa2001} IMF.	The parameter grid of the \texttt{M11-MILES} models ranges between 50 ages from $6.5\,\mathrm{Myr}$ to $15\,\mathrm{Gyr}$ and 10 metallicities [Z/H] from $-2.25$ to $0.35$. At $[Z/H]=-1.3$ and $[Z/H]=-2.3$, the age coverage is limited to $>2\,\mathrm{Gyr}$ and $>5\,\mathrm{Gyr}$, respectively (due to the stellar age distribution of the input library). The exact grid coverage is shown in Fig. \ref{fig:grid}.
	
	The second variant uses an updated version of the \texttt{MaStar SSP} models described in \citet{Maraston2020}, which are based on the MaStar stellar library \citep{Yan2019}. One of the main advantages to use the \texttt{MaStar SSPs} for fitting MaNGA data is that the empirical spectra are taken with the exact same instrument, hence, they match the galaxy observations in wavelength coverage ($3600\,$\AA$\,$ to $10300\,$\AA) and resolution (R=1400 to R=2200). The models are based on the same population synthesis code and input physics as \texttt{M11-MILES} \citep{Maraston2011}, described in detail in \citet{Maraston2005}. For the DR17 version of the \texttt{MaNGA {\sc firefly} VAC} we employ version 1.1 of the models -- dubbed `gold' -- which are based on the nineth MaStar Product Launch (MPL-9) of the  MaStar stellar library\footnote{\url{http://www.icg.port.ac.uk/mastar}}. These models use a combination of the stellar parameters derived by \citet{Hill2021}, \citet{Chen2020} and Lazarz et al. (in prep.). The `gold' version used here represents the best-performing ones among them, according to the calibration criteria described in \citet{Maraston2020}. The gold MPL-9 model library includes templates down to $3\,\mathrm{Myr}$. The full grid covers ages from $3\,\mathrm{Myr}$ to $15\,\mathrm{Gyr}$ and metallicities [Z/H] from $-2.25$ to $0.35$. 
We note that the current \texttt{MaStar} version does not include yet the thermally pulsating asymptotic giant branch (TP-AGB) contribution to the stellar evolution. The \texttt{MaStar} population models using MPL-11 and the further set of parameters by \citet{Imig2021} will be published in Maraston et al. (2022, in prep.). In addition to age and metallicity, a third parameter of the \texttt{MaStar} template grid allows for a flexible low-mass IMF slope from s=0.3 to 3.8. We adopt here a \citet{Kroupa2001} IMF with slope s=1.3 for consistency with the other model variant and previous studies, and leave the IMF analysis to future papers. As can be seen in Fig. \ref{fig:grid}, the parameter coverage of \texttt{MaStar} is much improved as compared to \texttt{M11-MILES} thanks to the significantly larger sample size of the MaStar stellar library, which extends to hotter temperatures leading to a larger parameter range in models. To summarise, the full wavelength range, the same instrument and the excellent parameter coverage make the \texttt{MaStar SSPs} a very suitable library to fit MaNGA observations.
	
\section{Workflow}
\label{sect:workflow}

	The main task to construct the \texttt{MaNGA {\sc firefly} VAC} is full spectral fitting of Voronoi-binned MaNGA data. This is the first out of two steps and it is executed employing the {\sc firefly} code for each binned spectrum in each datacubes separately, in other words, $\sim 3.7$ million times in total. The second step is performed by a single {\sc python} script that collects all output, calculates global galaxy parameters and produces the final \texttt{VAC} file.
	
\begin{figure*}
			\centering
			\includegraphics[trim=0cm 12cm 0cm 4.5cm, clip=True, width=15cm]{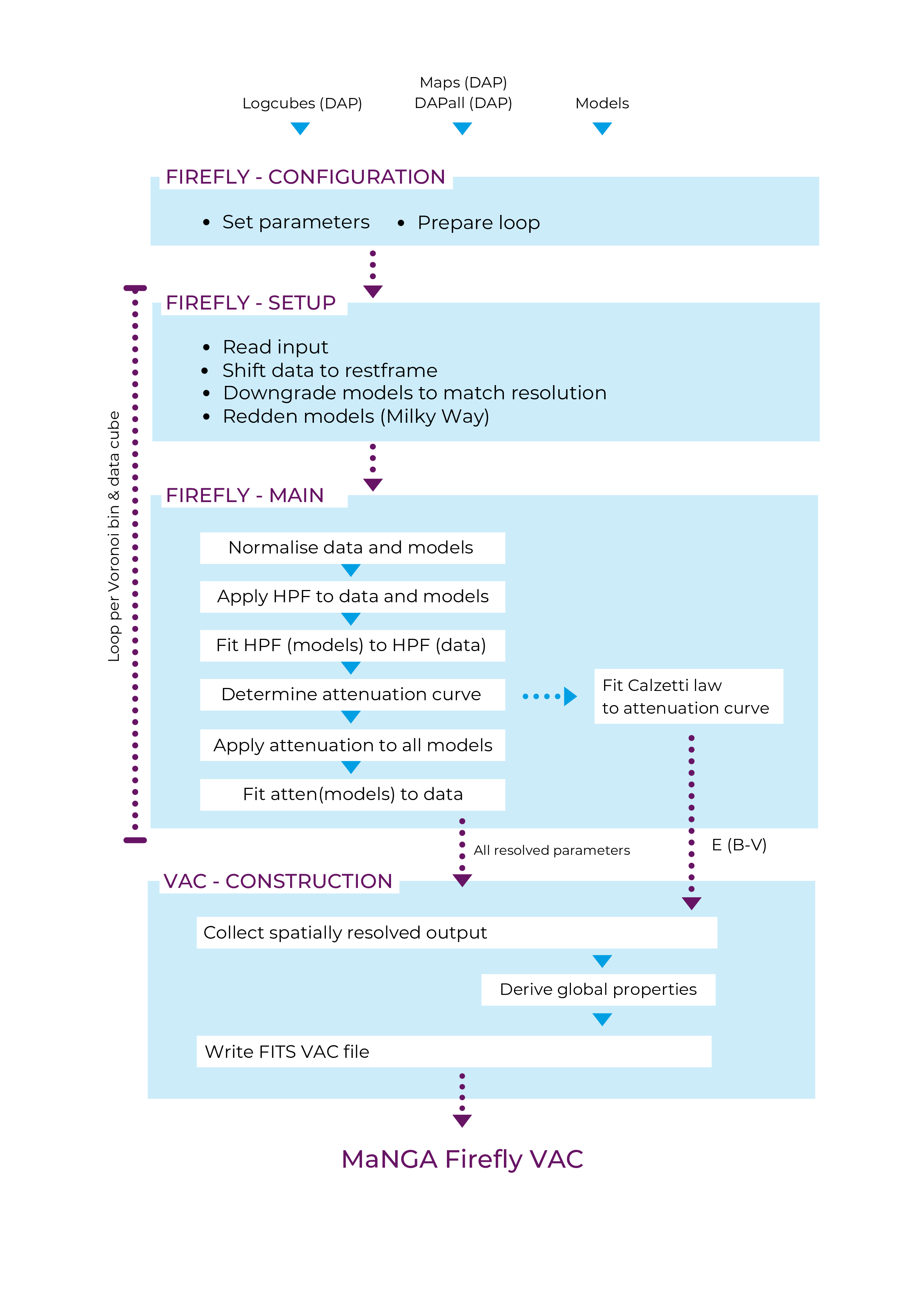}
    		\caption{Workflow. Details are described in Sect. \ref{sect:workflow}.}
    		\label{fig:workflow}
		\end{figure*}

	In the following, we first give a short general summary of {\sc firefly}, then, we present the specific setup and the run used for this catalogue, and lastly, we describe the final construction of the VAC. The workflow is summarised in Fig. \ref{fig:workflow}.
	
	\subsection{Firefly: code}
	
	{\sc firefly}\footnote{\url{https://www.icg.port.ac.uk/firefly/}} \citep{Wilkinson2017} is a full spectral fitting code written in {\sc python} designed to obtain properties of stellar populations from spectral integrated light observations. It compares arbitrarily weighted linear combinations of single-burst models to the observed spectrum and aims at iteratively minimising $\chi^2$ controlled by the Bayesian Information Criterion. All solutions within a statistical cut are retained, which allows to obtain uncertainty intervals based on the likelihood distribution of fitting solutions.
	
	The code was written with low S/N surveys, such as BOSS \citep{Eisenstein2011,Dawson2013} and MaNGA \citep{Bundy2015}, in mind and was shown to perform well down to S/N$\sim5$ \citep[Figs. 8-16 in][]{Wilkinson2017}. One of the main concepts is to allow for a large fitting freedom and sufficient exploration of the parameter space. As such, neither additive nor multiplicative polynomials are employed and star formation histories are not regularised. A novel method is used to account for dust attenuation, in which a high-pass filter (HPF) is applied to the spectrum to isolate small-scale features, such as absorption lines, from large-scale variations, such as dust attenuation but also inaccurate flux calibration. For more detail about the code as well as extensive performance tests, we refer to \citet{Wilkinson2017}. Further applications and testing of performance are presented in \citet{Goddard2017} and \citet{Comparat2017}.

		\subsubsection{Updates of Firefly v1.0.1}
		
		A new version of {\sc firefly} has been released together with the \texttt{MaNGA {\sc firefly} VAC}: version 1.0.1. The update includes: (a) a major revision of the interface, (b) user-level changes/bug fixes and (c) fixes and changes to the DR17-specific MaNGA setup. A more detailed list can be found on the github webpage\footnote{\url{https://github.com/FireflySpectra/firefly_release}}, here we would like to point out a few essentials: A cap on the maximum number of fit objects to be created per iteration is removed, as in some cases it generated a bias in templates files used. \texttt{MaStar} and \texttt{M11-MILES-SG} SSPs are added. The MaNGA datacube-specific LSF \citet{Law2021} is used instead of a generic averaged MaNGA resolution. Functionality of emission line masking and masking of individual pixels is improved.

	\subsection{Firefly: configuration/setup}
	
	The {\sc firefly} configuration file contains a small number of important user-level parameters such as the stellar population library to be used, the IMF, limits on age and metallicity to constrain the library, whether or not to mask emission lines and which lines to mask, as well as which dust law to use to calculate the colour excess $E_{B-V}$. In addition, it includes some further technical parameters. The configuration of the DR17 \texttt{VAC} is presented in Table \ref{tbl:vac_config}.
	
	The setup procedure prepares the input data (from MAPS, LOGCUBE and DAPall) before the main {\sc firefly} run can be executed. This is done for each spectrum separately and, thus, it is part of the loop over all Voronoi-binned spectra from all \texttt{DAP} datacubes. First, the observed spectrum and the emission line spectrum is read and the latter subtracted from the former.  The bitmask and inverse variance arrays are passed on to {\sc firefly}. Second, redshift and relative stellar velocity are used to shift the emission line-subtracted spectrum to rest frame wavelengths. Third, all model templates are downgraded by matching the intrinsic model resolution to the LSF and stellar velocity dispersion. Furthermore, all models are reddened \citep{Fitzpatrick1999} to match the line-of-sight Milky Way reddening of the observations using the coordinates of the source and the maps of \citet{Schlegel1998}.

	\begin{table}
	\caption{{\sc firefly} configuration for the DR17 VAC}
	\label{tbl:vac_config}
	\centering
	\begin{tabular}{lll} % four columns, alignment for each
		\hline
			& \texttt{FF-Mi} & \texttt{FF-Ma} \\
		\hline
		models & m11-MILES & MaStar-gold \\
		IMF & Kroupa & Kroupa \\
		age limits & $<$ `AoU' (z) & $<$ `AoU' (z) \\
		Z limits & none & none \\
		dust law & Calzetti+2000 & Calzetti+2000 \\
		em-line masking & off & off \\
		\hline
	\end{tabular}
	\medskip\\
	\raggedright
	Notes: \texttt{FF-Mi} and \texttt{FF-Ma} refer to the two variants of the DR17 \texttt{MaNGA {\sc firefly} VAC} employing the \texttt{m11-MILES} and the \texttt{MaStar-gold} model libraries, respectively (see Sect. \ref{sect:ssps}). `AoU'(z) means the age of the universe at the respective redshift. Calzetti+2000 refers to the dust law presented in \citet{Calzetti2000}.
	\end{table}	
	
	\subsection{Firefly: run}
	\label{sect:ff_run}
	
		\begin{figure*}
			\centering
			\includegraphics[width=18cm]{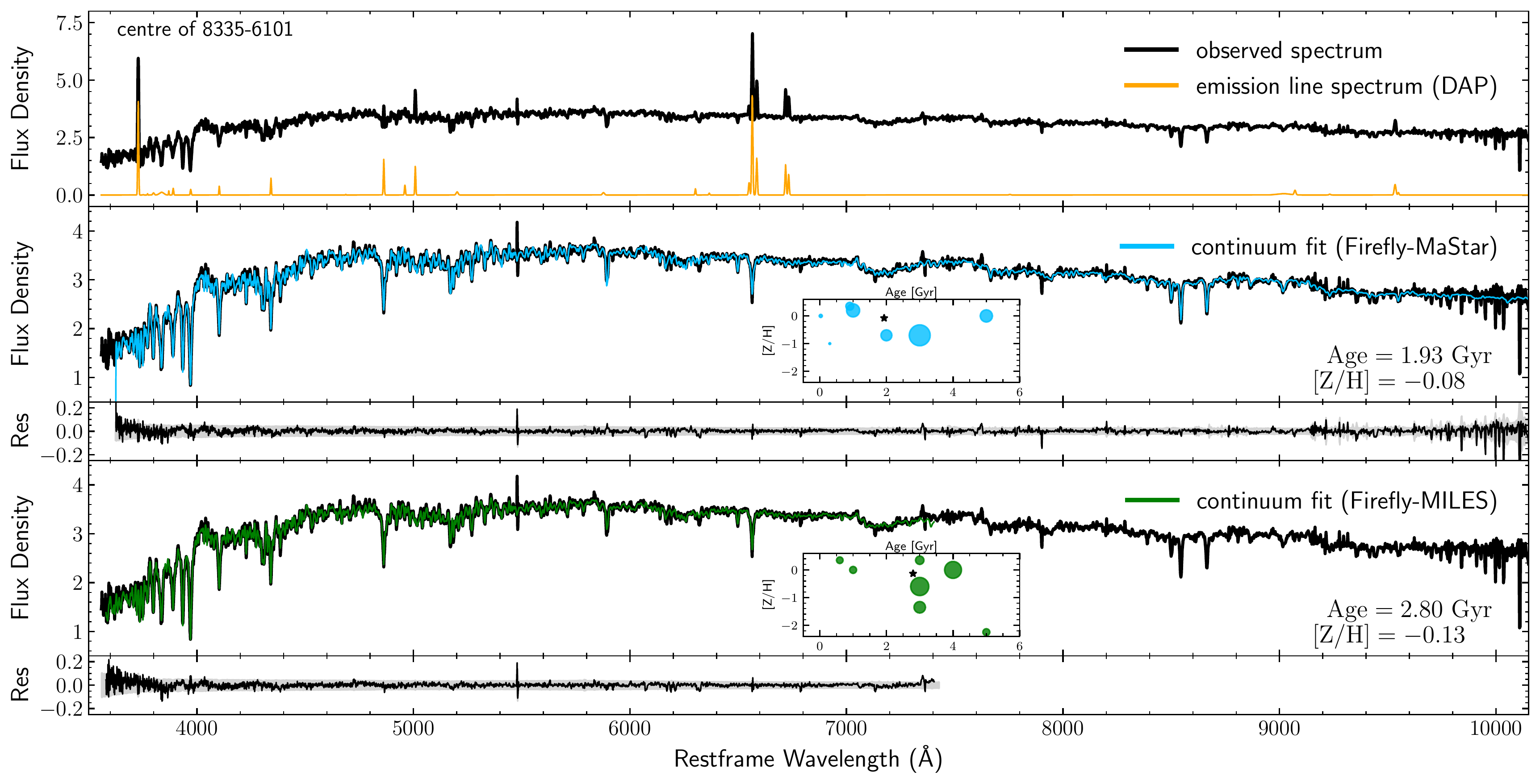}
    		\caption{Example spectrum and modelling from the central spaxel of \texttt{PLATEIFU}=8335-6101 ($z=0.018$). \textit{Top:} Observed spectrum and emission line spectrum as derived by the {\sc dap}. \textit{Middle:} Emission line subtracted spectrum and stellar continuum fit by {\sc firefly} using the \texttt{MaStar} SSPs. Residuals of the fit are shown below the spectrum. The grey shaded area shows the uncertainty on the observation. Inset plot shows the star formation history, i.e. the age-metallicity grid of SSPs used to fit the spectrum. The size of the dots corresponds to the mass fraction contributed to the total population. Mean age and metallicities are shown as star symbols and are annotated in the bottom right. \textit{Bottom:} Same as middle panel but employing the \texttt{M11-MILES} SSPs.}
    		\label{fig:spec1}
		\end{figure*}
	
	The main steps of the {\sc firefly} fitting procedure are outlined in Fig. \ref{fig:workflow}; a slightly more detailed version of this part of the workflow is shown in \citet[][Fig. 3]{Wilkinson2017}.
	
	(1) The observed spectrum and all model templates are normalised to the total integrated flux. The normalisation factors are saved and will be used later to convert light-weights to mass-weights. (2) An HPF is applied to both data and models to remove long wavelength modes. (3) A weighted linear combination of the filtered model templates are fitted to the filtered data. (4a) The attenuation curve is determined as
	
	\begin{equation}
	F_{\mathrm{Atte}}(\lambda) = \frac{F_{\rm data}(\lambda)}{F_{\rm bestfit}(\lambda)} - \frac{\mathrm{HPF}(F_{\rm data}(\lambda))}{\mathrm{HPF}(F_{\rm bestfit}(\lambda))}.
	\end{equation}
	
	In the first term of that equation, $F_{\rm data}(\lambda)$ is the unfiltered data spectrum and $F_{\rm bestfit}(\lambda)$ is the reconstructed full best fit model combination. The ratio between both is the sum of the large-scale variations (i.e. the attenuation curve) and the residuals of the fit. The latter must therefore be subtracted to obtain $F_{\mathrm{Atte}}(\lambda)$. Subsequently, the attenuation curve is smoothed. (4b) In a parallel step, the attenuation curve is fitted with the \citet{Calzetti2000} law to obtain the colour excess $E_{B-V}$ (shown on the right-hand side of Fig. \ref{fig:workflow}). (5) The smooth attenuation curve (not the Calzetti-fitted curve) is applied to all unfiltered model templates. (6) The unfiltered data is fitted in a second fitting cycle with a weighted linear combination of the reddened unfiltered model templates. All solutions within a statistical cut are retained to construct the likelihood distribution. The best fit is the peak of the distribution and the confidence intervals for each stellar population parameter are obtained within the corresponding likelihood intervals \citep[see Fig. 7 in][]{Wilkinson2017}.
	
\begin{figure*}
			\centering
			\includegraphics[width=18cm]{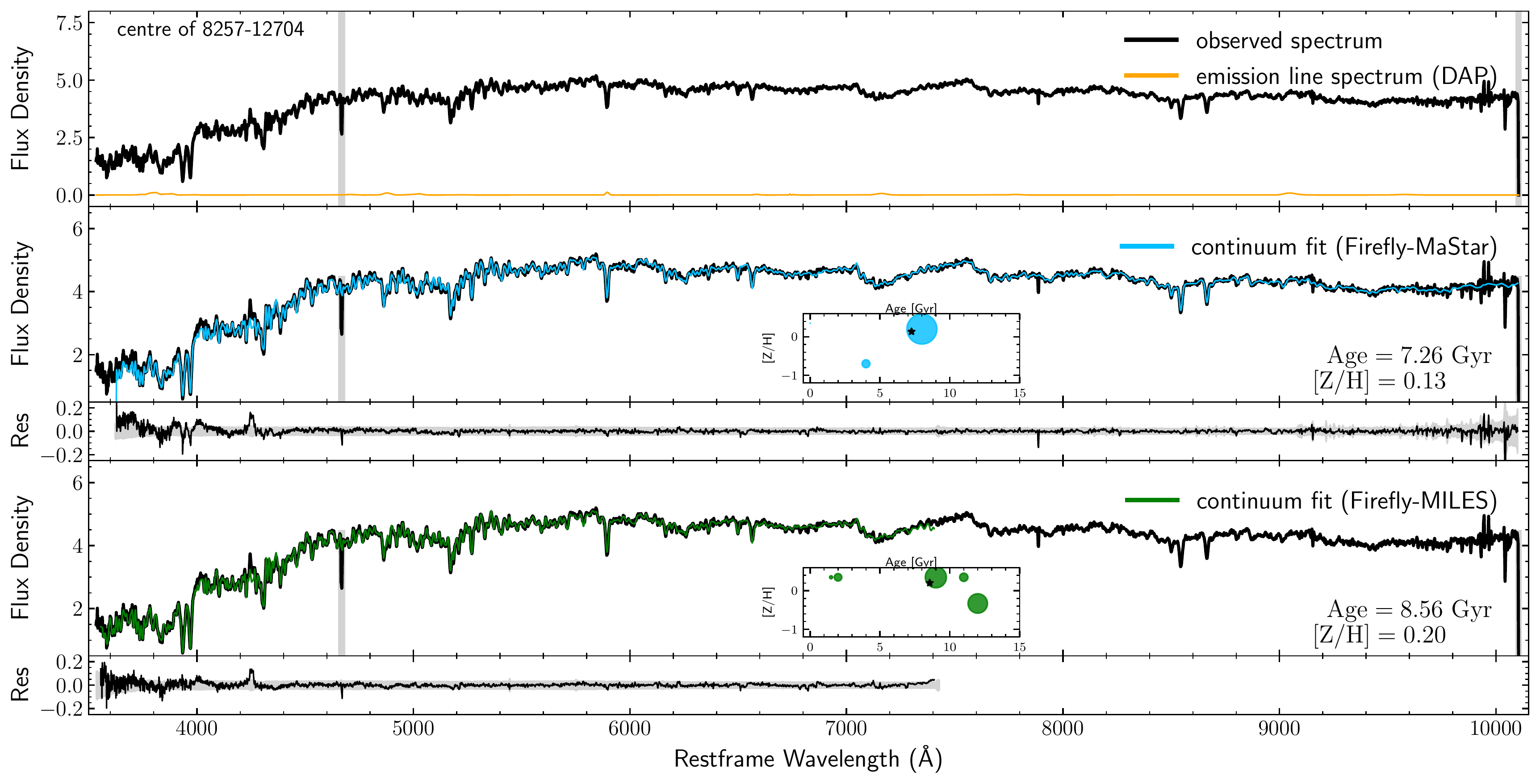}
    		\caption{Same as Fig. \ref{fig:spec1} but for a spectrum from the central spaxel of \texttt{PLATEIFU}=8257-12704 ($z=0.025$) showing no emission lines and an old stellar population. Grey shaded wavelength regions are masked during fitting.}
    		\label{fig:spec2}
		\end{figure*}

	The output of the {\sc firefly} run consists of the observed input spectrum, the associated error spectrum, the best fit model spectrum, and a series of stellar population parameters: the full SFH with light-weights and mass-weights of the individual SSPs, light- and mass-weighted average age and metallicity defined as
	
	\begin{equation}
	\langle\mathcal{P}\rangle_{\rm LW} = \sum_{i=1}^n w_i^{\mathrm{L}} \mathcal{P}_i, \qquad \langle\mathcal{P}\rangle_{\rm MW} = \sum_{i=1}^n w_i^{\mathrm{M}} \mathcal{P}_i,
	\end{equation}
	
	where $\mathcal{P}$ is either age or $Z$, $w_i^{\mathrm{L}}$ is the light-weight and $w_i^{\mathrm{M}}$ the mass-weight of the $i$-th SSP. Note that these parameters are linearly averaged. Further outputs are the colour excess $E_{B-V}$, the total stellar mass and its partition into masses of living stars, white dwarfs, neutron stars, black holes and stellar ejecta. The division of mass is obtained by applying mass loss factors on the mass of each SSP contribution that depend on age, metallicity and IMF. The adopted initial mass-final mass theoretical relations are described in \citet{Maraston1998}, \citet{Maraston2005} and are based on \citet{Renzini1993}. Two examples of a fit with {\sc firefly} to the emission line subtracted stellar continuum of the central spaxel are shown in Fig. \ref{fig:spec1} and Fig. \ref{fig:spec2}, for a relatively young population with strong emission lines and for a quiescent old population, respectively. The figure also exemplifies the slightly different results obtained according to the choice of the model library.

	\subsection{VAC construction}
	
	After fitting all individual spectra of all datacubes with {\sc firefly}, the \texttt{VAC construction {\sc python} script} is executed. The main purpose of this script is to (1) collect all output from {\sc firefly}, (2) derive global stellar population properties (see next subsections), and (3) write everything in one single FITS file. The only spatially resolved property, in addition to the {\sc firefly} output, that is computed as part of this script is the stellar surface mass density, which is derived as
	
	\begin{equation}
	\Sigma_{\star, i} = M_{\star, i} / A_i,
	\end{equation}
	
	where $M_{\star, i}$ is the stellar mass and $A_i$ is the projected\footnote{We do not apply an inclination correction.} surface area of the $i$-th Voronoi bin in any given galaxy.
	
		\subsubsection{Central values and values at 1 $R_\mathrm{e}$}
		In addition to spatially resolved stellar population measurements, as part of the \texttt{MaNGA Firefly VAC}, we also provide global galaxy properties, specifically, characteristic values such as the average age and metallicity in the galaxy centre, at $1\,R_\mathrm{e}$, and gradients within $1.5\,R_\mathrm{e}$.  		
		
		We measure the central average stellar population properties within a fixed circular aperture of $3\arcsec$ diameter. The motivation of this choice is to provide a direct comparison from the IFU observations of MaNGA galaxies to large SDSS catalogues from single-$3\arcsec$-fibre spectroscopy. The obvious caveat is that these measurements are not sensitive to inclination and distance effects. In case the user of this \texttt{VAC} is interested in central values that are averaged over a more similar physical space in each galaxy, we advise to derive those from the spatially resolved measurements in an elliptical aperture of flexible size.
		
		In contrast to the central values, we employ elliptical apertures for the derivation of the characteristic values at the effective radius. In particular, the value at the effective radius is averaged within two concentric ellipses at $0.9\,R_\mathrm{e}$ and $1.1\,R_\mathrm{e}$, while having the angle of the semi-major axis aligned with the position angle of the galaxy. 
		
		It is important to point out that we do not rebin and reanalyse the spectra but, in fact, we average the results from the Voronoi-binned {\sc firefly} analysis. In detail, the average of the light-weighted (mass-weighted) parameters is calculated as the mean of all Voronoi bins within the aperture weighted by their relative flux (mass) contribution:
		
		\begin{equation*}
			\langle\mathcal{P}\rangle_\mathrm{LW} = \frac{\sum_{i=1}^n \mathcal{P}_{\mathrm{LW}, i} f_i a_i}{\sum_{i=1}^n f_i a_i}, \quad \langle\mathcal{P}\rangle_\mathrm{MW} = \frac{\sum_{i=1}^n \mathcal{P}_{\mathrm{MW}, i} m_i a_i}{\sum_{i=1}^n m_i a_i},
		\end{equation*}
		
		where $n$ is the total number of Voronoi bins within the aperture. For the $i$th Voronoi bin, $\mathcal{P}_i$ is the measured parameter, $f_i$ is the $g$-band-weighted mean flux, $m_i$ is the stellar mass as returned by {\sc firefly}, and $a_i$ is the fraction of the bin area that is inside the aperture. These values are always computed as long as the number of included Voronoi bins is non-zero. Errors are calculated from the weighted average upper and lower error boundaries of the $\mathcal{P}_i$.
		Light- and mass-weighted age and metallicity within the central $3\arcsec$ and at $1\,R_\mathrm{e}$ are provided in the FITS extension \texttt{HDU2}.		
		
		\subsubsection{Gradients within 1.5 $R_\mathrm{e}$}
		\label{sect:gradients}
		
		Similar to the average value at the effective radius, we also derive the radial gradients of each galaxy in elliptical coordinates (i.e. taking into account the inclination), but within a larger aperture of $1.5\,R_\mathrm{e}$. We impose a minimum number of ten bins within the aperture as a requirement to compute the gradient.
		
		The gradient is calculated by first running a median along ten equally-sized radial bins and subsequently fit a linear regression to the binned median values. We estimate the statistical uncertainty of each gradient by performing 100 bootstrap realisations of the linear fit using randomly resampled bins. The light- and mass-weighted age and metallicity gradients as well as the corresponding zeropoints are given in the FITS extension \texttt{HDU3}.
		
		\subsubsection{Changes compared to DR15 VAC}
		
		The major change in the \texttt{MaNGA {\sc firefly} VAC} is the addition of stellar population properties calculated with the new SDSS-IV-based MaStar set of stellar population models, covering the full wavelength range of the data. In addition, we decided to drop absorption line strength indices in DR17, which were included in SDSS DR15 and earlier versions of the \texttt{VAC} \citep[c.f.][]{Goddard2017}, as these are now provided by the {\sc dap}.		
		
		Further small changes to the DR15 version include that the radius in \texttt{HDU4} is now given in elliptical coordinates and the azimuth is added. Masses in \texttt{HDU11} and \texttt{HDU12} are given per spaxel and as total mass per Voronoi cell. 
		
		\subsubsection{Star formation rates}
		\label{sect:sfr}
		
		The latest addition to the \texttt{MaNGA {\sc firefly} VAC} is the derivation of spatially resolved and global star formation rates (SFRs). This has been performed after the SDSS VAC release and is, therefore, not part of the DR17 version of the VAC. We will make the full table of SFRs publicly available as separate files and as part of upcoming versions of the VAC.
		
		SFRs are calculated directly from the derived stellar masses and ages by integrating mass fractions over a given age range. The range to be considered is somewhat arbitrary and depends on the definition of \textit{current} SFR. For example, the last 10 Myr trace the time interval of photoionisation from young, massive stars probed by the H$\alpha$ recombination line, while 100 Myr is often used to calibrate star formation tracers from the far infrared \citep[e.g.][]{Kennicutt1998}. We provide both measurements and define the SFR of the $i$-th Voronoi bin as:
		
		\begin{equation}
		\mathrm{SFR}_i = \left. \sum_{\mathrm{SSP}_{t=0}}^{t_n} M_{\mathrm{SSP},i} \middle/ \Delta t \right.,
		\end{equation}

		where $M$ is the total mass of a certain SSP template (not corrected for mass loss) and the sum goes over all SSP templates of age $t=0\,\mathrm{Myr}$ to $t_n = 10\,\mathrm{Myr}$ or $100\,\mathrm{Myr}$. The total SFR per galaxy is then derived as sum over all local SFRs across the galaxy FoV.

\section{Data products}
\label{sect:output}

	\subsection{Overview}
	
	The complete \texttt{VAC} is provided in a single {\sc fits} file per model variant \texttt{manga-firefly-v3\_1\_1-miles.fits} and \texttt{manga-firefly-v3\_1\_1-mastar.fits} of $\sim 6.1\,\mathrm{GB}$ size and include all 16 Header-Data Units (\texttt{HDU}s) as detailed below. Additionally, there is a light-weight version of only global parameters \texttt{manga-firefly-globalprop-v3\_1\_1-miles.fits} and \texttt{manga-firefly-globalprop-v3\_1\_1-mastar.fits} of $\sim 2.8\,\mathrm{MB}$ size containing \texttt{HDU0}--\texttt{HDU3}. All four files can be accessed from the SDSS webpage.\footnote{\url{https://data.sdss.org/sas/dr17/manga/spectro/firefly/v3_1_1}}
	
	\begin{itemize}
		\item \texttt{HDU0}: empty
		\item \texttt{HDU1}: general galaxy information, pipeline versions
		\item \texttt{HDU2}--\texttt{HDU3}: global galaxy stellar population properties
		\item \texttt{HDU4}--\texttt{HDU15}: spatially resolved parameters
	\end{itemize}
	
	A detailed description of the {\sc fits} file content is given in Table \ref{tbl:vac_content} and in the official datamodel.\footnote{\url{https://data.sdss.org/datamodel/files/MANGA_FIREFLY/FIREFLY_VER/manga_firefly.html}} In addition, we provide spatially resolved and global star formation rates, as well as the results from fitting-performance analyses in separate files on the ICG institute's website.\footnote{\url{http://www.icg.port.ac.uk/manga-firefly-vac/}}

	\subsection{FF-Mi versus FF-Ma}
	
		\begin{figure*}
			\centering
			\includegraphics[width=18cm]{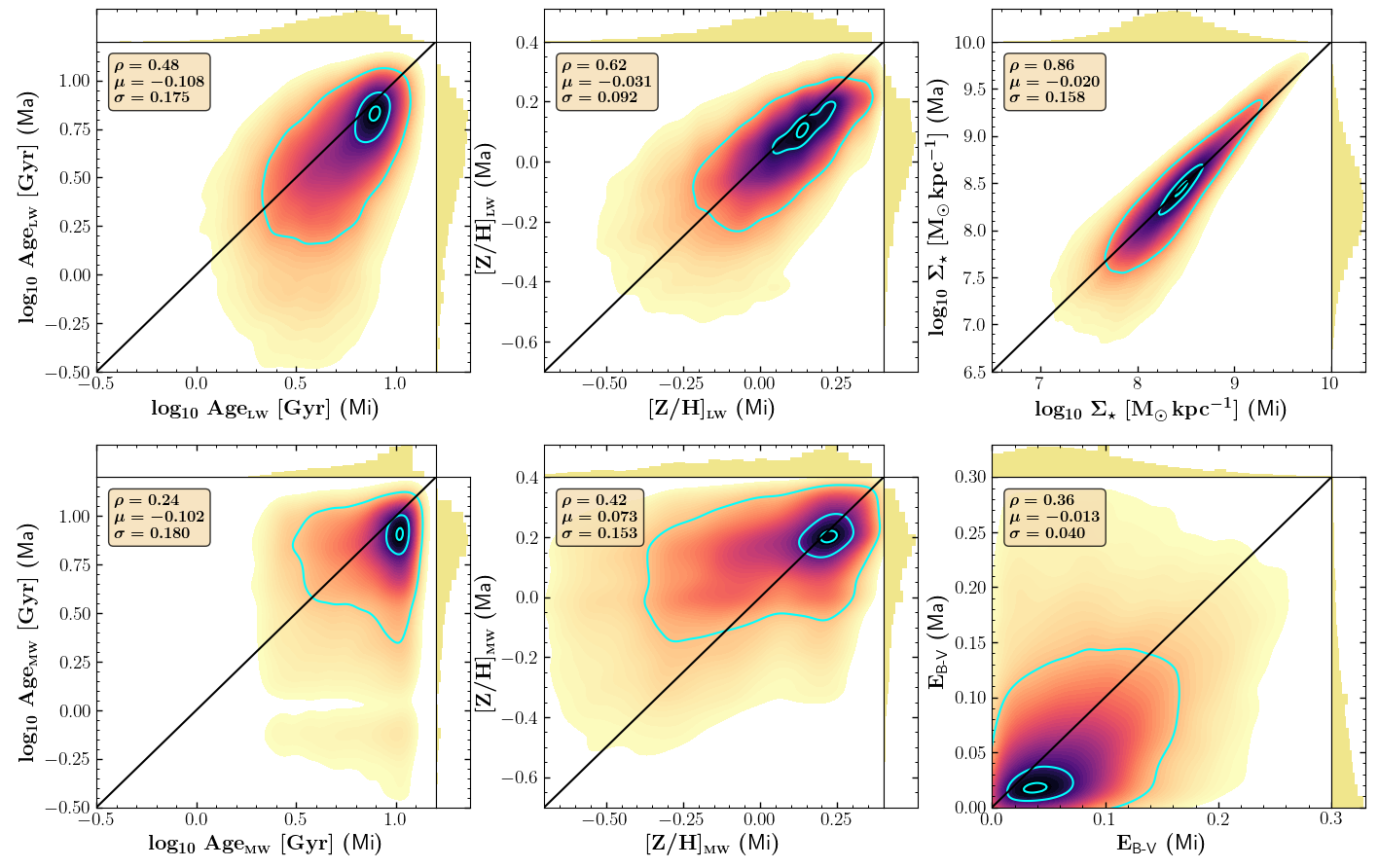}
    		\caption{Comparison of the stellar population properties of MaNGA galaxies between both \texttt{VAC} variants \texttt{FF-Mi} and \texttt{FF-Ma}. Density plots visualise the 2D distribution while histograms on top and to the right of each panel show the 1D distribution of each property. Colours and histograms are in linear scales where darker colours correspond to higher densities. Cyan contours enclose $1\%$, $10\%$ and $68\%$ of the distribution. In the top left corner we annotate the Spearman's rank correlation coefficient $\rho$, the median difference $\Delta \mu$ between the parameters $y-x$ and the median absolute deviation $\sigma$.}
    		\label{fig:mi_vs_ma}
		\end{figure*}
	
	\subsubsection{Individual fits}	
	
	In this section, we show the distributions of the derived parameters and at the same time compare both variants of the catalogue. In Figs. \ref{fig:spec1} and \ref{fig:spec2}, we illustrate how the use of different model libraries can lead to different best fit linear combinations of SSPs and, thus, different average quantities. In Fig. \ref{fig:mi_vs_ma}, we compare the spatially resolved average stellar population parameters for all fits with $S/N>5$ in the \texttt{VAC} ($\sim 3.7$ million) between the \texttt{FF-Mi} and the \texttt{FF-Ma} variants.
	
	The best agreement is found for stellar masses, here represented as surface mass density; not surprisingly, as this is known to be typically the most robust parameter. Note, however, that different choices of IMFs, isochrones and spectral fitting codes lead to systematic offsets that we discuss in Sect. \ref{sect:masses}.
	
	Looking at light-weighted metallicity and age, we find that metallicity agrees reasonably well with an average difference of $\Delta \mu_\mathrm{[Z/H],LW}\equiv \mathrm{[Z/H]_{LW,Ma}-[Z/H]_{LW,Mi}} = -0.03 \pm 0.09\,\mathrm{dex}$. A larger systematic difference is seen between the ages with $\Delta \mu_\mathrm{Age, LW} = -0.11 \pm 0.18$. \texttt{MILES}-derived mean ages are on average older than \texttt{MaStar}-derived ages at all age bins. In particular, \texttt{FF-Mi} does not go younger than 1\,Gyr. \citet{Maraston2011} already noted that \texttt{MILES}-based population models led to older ages with respect to to e.g. population models based on \texttt{STELIB}. Part of the reason might be intrinsic in the stellar parameters associated to the \texttt{MILES} library. Part of the effect, however, simply comes from the parameter coverage in that the \texttt{M11-MILES} library lacks low-metallicity low-age templates. When such stellar populations occur, these absent templates are likely to be replaced by the next possible older ones leading to older ages on average. Following this scenario, we are also able to explain the higher dust attenuation derived in \texttt{FF-Mi}. In the first fitting loop, {\sc firefly} determines the best fit template combination based only on the small scale variations in the spectrum. An older best fit SSP combination will lead to smaller fluxes when the full spectrum is compared to the data. This will subsequently be compensated for by the attenuation curve (known as age-dust degeneracy), leading to higher attenuation values in \texttt{FF-Mi} as compared to \texttt{FF-Ma}.
	
	Light-weighted averages are dominated by the young stellar populations and it is, thus, on the one hand, not surprising that mass-weighted ages are significantly shifted towards older average ages in both variants of the VAC. The density plot draws the attention to a tail of some populations that appear old in \texttt{FF-Mi} but quite young for mass-weighted ages in \texttt{FF-Ma}. However, this is a very small fraction of the data and the majority of data points are confined to old average ages on both axes with a mean difference that remains, in fact, almost unchanged as compared to the light-weighted counterparts. The distributions of ages are in line with our current understanding that more than half of the stellar mass of galaxies has already been in place 10\,Gyr ago \cite{Madau2014}. Explaining differences between light- and mass-weighted metallicites, on the other hand, is not straight forward. The mass-weighted metallicities in \texttt{FF-Ma} are on average $\Delta \mu_\mathrm{[Z/H],MW} = 0.07 \pm 0.15\,\mathrm{dex}$ higher than in \texttt{FF-Mi}. This is possibly caused by the age-metallicity degeneracy in that the older SSPs are compensated by more metal-poor SSPs. The difference is not seen in light-weighted averages and is even slightly reversed. The aforementioned compensation of young templates can happen dominantly in the old component and is even more likely to happen in the old component given the lack of young, metal-poor templates. This explains why the metallicity difference between \texttt{FF-Mi} and \texttt{FF-Ma} is only seen in the mass-weighted plot. In fact, the near-infrared extension of \texttt{MaStar} is able to capture more metal lines that are more relevant in the older populations and, therefore, metallicity should be better constrained in \texttt{FF-Ma}.
	
	\begin{table*}
	\caption{{\sc firefly} setup for the configuration tests}
	\label{tbl:vac_test_config}
	\centering
	\begin{tabular}{llllllll} 
		\hline
			& \texttt{FF-Mi} & \texttt{FF-Ma} & config1 & config2 & config3 & config4 & config5 \\
		\hline
		models & M11-MILES & MaStar-gold & MaStar-gold & M11-MILES & M11-MILES-SG & MaStar-gold & MaStar-gold\\
		IMF & Kroupa & Kroupa & Kroupa & Kroupa & Kroupa & Salpeter & Kroupa\\
		wavelength range & short & full & full & short & short & full & short\\
		em-line masking & off & off & on & on & off & off & off\\
		\hline
	\end{tabular}
	\end{table*}

	To shed further light on differences between light- and mass-weighted metallicity averages, it is informative to compare them directly. From a close-box chemical evolution scenario, one would expect younger stellar populations to be more chemically mature or in other words more metal-rich. Hence, the light-weighted average population would have a higher metallicity than the mass-weighted one. Yet, at a given age, the lower the metallicity the brighter the spectrum; a circumstance acting contrary to the effect of the age. In addition to both of that, chemical evolution of galaxies is not as simple as a closed-box scenario and inflows, outflows, feedback and mergers make these simplistic views more complex.
	
\begin{figure}
			\centering
			\includegraphics[width=\columnwidth]{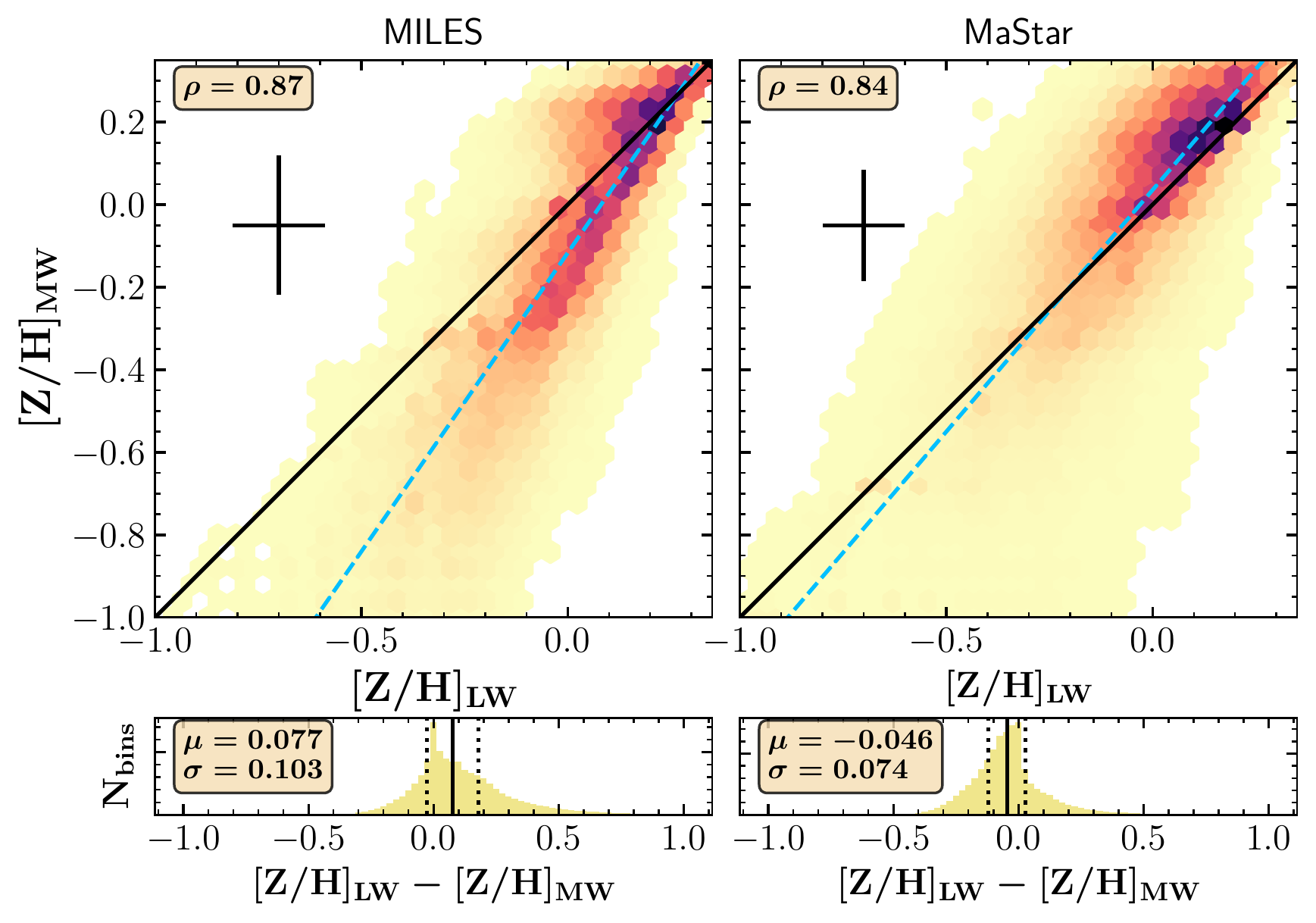}
    		\caption{Comparison between light-weighted and mass-weighted average metallicity for both \texttt{VAC} variants. The errorbar shows the median individual uncertainty of each measurement. A linear regression is shown as blue dashed line. The parameters $\rho$, $\mu$ and $\sigma$ have the same meaning as in Fig. \ref{fig:mi_vs_ma}.}  
    		\label{fig:lw-mw}
		\end{figure}

	In Fig. \ref{fig:lw-mw}, we directly compare mass-weighted to light-weighted metallicity for both \texttt{VAC} variants. Both parameters are strongly linearly correlated. Yet, the slope of the correlation is larger than one yielding a crossing of the one-to-one line such that at low metallicities  mass-weighted values are lower while at high metallicities light-weighted values are lower. In \texttt{FF-Mi}, lower mass-weighted values dominate most of the distribution, while in \texttt{FF-Ma}, mass-weighted values are more often higher. The differences between both variants probably arise from the covered wavelength range. In \texttt{FF-Mi} bluer wavelengths are covered, the flux of which is dominated by younger populations yielding higher light-weighted metallicities. With \texttt{FF-Ma} fitting redder wavelengths, this effect becomes less dominant and even slightly reversed at high metallicities.
	
	\subsubsection{Gradients}
	
\begin{figure}
			\centering
			\includegraphics[width=\columnwidth]{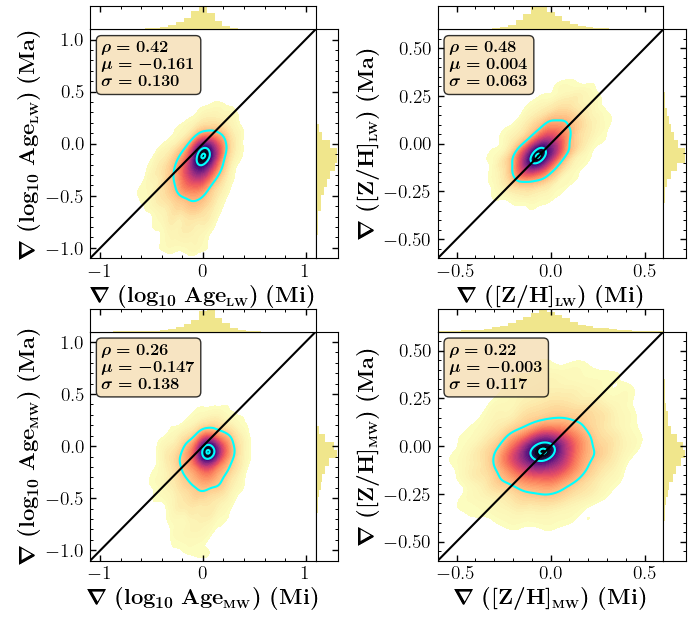}
	    	\caption{Comparison of linear radial age and metallicity gradients derived within 1.5$\,R_\mathrm{e}$ between both \texttt{VAC} variants. Colour scale, contours and statistical parameters have the same meaning as in Fig. \ref{fig:mi_vs_ma}.} 
    		\label{fig:gradients}
		\end{figure}

	Radial gradients of gas-phase abundances or stellar populations in galaxies have been extensively studied in the literature \citep[e.g.][]{Sanchez-Blazquez2014,GonzalezDelgado2015,Belfiore2017,Goddard2017,
	Poetrodjojo2018,Lian2018,Oyarzun2019,Lacerna2020,Neumann2020, Neumann2021}. The \texttt{MaNGA {\sc firefly} VAC} calculates gradients on an individual galaxy-by-galaxy basis as described in Sect. \ref{sect:gradients}. It is instructive to study how the individual differences between \texttt{FF-Mi} and \texttt{FF-Ma} fits affect global galaxy properties such as gradients. In Fig. \ref{fig:gradients} we present and compare the distribution of light-weighted and mass-weighted age and metallicity gradients for both VAC variants.
	
	Light-weighted gradients are well correlated with an excellent agreement between the metallicity gradients. Age gradients are systematically more negative in \texttt{FF-Ma}. This indicates that the age difference for individual fits as seen in Fig. \ref{fig:mi_vs_ma} is larger at larger radii, which are often dominated by young star-forming regions. Mass-weighted gradients are very flat in both variants. Age and metallicity peak close to zero in both distributions. Age gradients are again slightly smaller in \texttt{FF-Ma}. The scatter in the metallicity gradient distribution is on the order of the measurement error (see e.g. Fig. \ref{fig:lw-mw}) and likely reflects the uncertainty in determining mass-weighted gradients.
		
	\subsection{Differential effects of input parameters}
	\label{sect:config_tests}
	
		To further aid our understanding of the differences between both variants of the \texttt{VAC} and to test the robustness of the results, in this subsection, we explore the effect of the choice of the IMF, the fitted wavelength range, the model parameter grid and emission line masking on the derived stellar population parameters.
		
		For these tests, we semi-randomly selected a subsample of 300 galaxies from the main MaNGA sample while monitoring that the distribution in stellar mass, colour, age and metallicity are representative of the overall sample. In addition to both \texttt{VAC} variants \texttt{FF-Mi} and \texttt{FF-Ma}, we created another five mini-VACs of these 300 galaxies with varying {\sc firefly} configurations. The setup is shown in Table \ref{tbl:vac_test_config}. Each mini-VAC contains $\sim 100,000$ sets of spatially resolved parameters. Detailed figures comparing the fitted parameters can be found in Appendix \ref{sect:config_plots}. Here, we discuss the main results.

		\subsubsection{Initial mass function}
		\label{sect:imf}
		
			Both main variants of the \texttt{VAC} use stellar population models that employ the \citet{Kroupa2001} IMF. While it is known that the choice of the IMF only affects the computed stellar mass \citep[e.g.][]{Pforr2012}, it is instructive to inspect all stellar population parameters and, at the same time to quantify the change in stellar mass. For this comparison, we use the exact same {\sc firefly} configuration with the \text{MaStar} model grid based on a low-mass IMF slope of 1.3 for a Kroupa IMF (\texttt{FF-Ma}) and a slope of 2.35 for a \citet{Salpeter1955} IMF (\texttt{config4}).
			
			Age, metallicity and dust attenuation show no systematic differences ($\Delta \mu \leq 0.01\,\mathrm{dex}$) with a scatter comparable to the average uncertainty of the measurements ($\Delta \sigma \leq 0.1\,\mathrm{dex}$). Surface mass density shows a clear systematic offset with small scatter towards higher masses when employing a Salpeter IMF. The median difference is $\Delta \mu = 0.170 \pm 0.075\,\mathrm{dex}$, in other words, models with a Salpeter IMF produce masses by a factor of 1.48 higher than models based on a Kroupa IMF in very good agreement with the theoretical factor of 1.5 reported in \citet{Maraston2005}, and slightly lower than the offset of 0.209\,dex found in \citet{Pace2019a}. From fitting broad-band photometry of mock galaxies over a wide wavelength range with a wide range of templates, \citet[][tables 3 and 4]{Pforr2012} report offsets of 0.28 for star-forming and 0.08 for quiescent galaxies at a redshift of z=0.5, on average in good agreement with our result. See also \citet{DominguezSanchez2019}, where they show that the mass-to-light ratio of MaNGA early-type galaxies is $\sim 1.5$ times higher for the Salpeter IMF than for the Kroupa (right panel of their Fig. 17).
			
		\subsubsection{M11-MILES: Squared grid models}
		
			One of the major differences between the \texttt{MaStar SSPs} and \texttt{M11-MILES} that we used to explain discrepancies between the \texttt{VAC} variants is the grid coverage, in particular the lack of low-metallicity low-age MILES templates. We try to test the effect by employing a special `squared-grid' version of the \texttt{M11-MILES} models, henceforth called \texttt{M11-MILES-SG}. These models use complementary high-resolution theoretical stellar population model spectra from \citet{Maraston2009} based on the \citet{RodriguezMerino2005} model atmospheres, then smoothed to the MILES resolution. The extended \texttt{M11-MILES-SG} models have been used in \citet{Trussler2020} for the analysis of DR17 SDSS integrated galaxy spectra.
			
			When comparing the results using \texttt{M11-MILES-SG} (\texttt{config3}) with \texttt{FF-Ma}, we find that the addition of young model templates does produce more similar light-weighted ages among both catalogues, especially in the lower age range. That also leads to a slight improvement for the mass-weighted metallicities. Yet, a disagreement clearly remains at older average ages. We conclude that template grid coverage is an important factor when comparing stellar population parameters, but it is not sufficient to explain the differences. On the other hand, it is also important to note that while \texttt{M11-MILES-SG} has an improved squared coverage, the sampling of the grid between the libraries remains non-identical.
			
		\subsubsection{Wavelength range}
		
			Another difference between MILES and MaStar is the wavelength coverage. We explore the importance of covering the full wavelength range in MaStar (i.e. $3600\,$\AA--$10300\,$\AA) by producing another mini-VAC using the exact same MaStar configuration except for the wavelength range that we limit to the range of the MILES library (\texttt{config5}).
			
			The fits with short wavelength range (i.e. 3600\,\AA--7430\,\AA) produce indeed on average slightly older ($\Delta \mu_\mathrm{Age,LW} = 0.04 \pm 0.17\,\mathrm{dex}$) and more metal-poor ($\Delta \mu_\mathrm{[Z/H],LW} = -0.05 \pm 0.09\,\mathrm{dex}$) results, an indication that the long MaStar/MaNGA wavelength range helps to break the age-metallicity degeneracy \citep[e.g.][]{Maraston2005}. Comparing the short wavelength range MaStar results with MILES, we find that the average differences are reduced but remain present.
			
			We further look at the combined effect of wavelength range and grid coverage by comparing the fits using the short MaStar models (\texttt{config5}) with those using the squared grid MILES models (\texttt{config3}). As expected, light-weighted age  ($\Delta \mu_\mathrm{Age,LW} = -0.03 \pm 0.17\,\mathrm{dex}$) and mass-weighted age  ($\Delta \mu_\mathrm{Age,MW} = -0.06 \pm 0.17\,\mathrm{dex}$) and metallicity ($\Delta \mu_\mathrm{[Z/H],MW} = 0.01 \pm 0.15\,\mathrm{dex}$) differences are reduced even further, yet remaining to a lower degree. We conclude that the differences between both main variants of our \texttt{MaNGA {\sc firefly} VAC} can be explained to a large extent by the combined effect of model template grid coverage and fitted wavelength range. The remaining discrepancies most likely lie in the assumed stellar parameters, which assign an empirical spectrum to a temperature-gravity-metallicity location of stellar evolution, investigated in detail in \citet{Maraston2020}.

		\subsubsection{Emission line masking}
		
			Finally, we compare fits with subtracted emission lines with fits with masked emission lines for both M11-MILES (\texttt{config2}) and MaStar models (\texttt{config1}). With both model libraries, we obtain on average slightly older and slightly more metal-rich parameters when emission lines are masked. However, differences are small ($\Delta \mu < 0.02\,\mathrm{dex}$). Since a lot of care has been taken to accurately model the emission lines in the {\sc dap} \citep{Belfiore2019}, we chose to use the emission line subtracted spectra without masking for both variants of the MaNGA Firefly VAC.

	\subsection{Stellar remnants}
	\label{sect:remnants}			

		\begin{figure}
			\centering
			\includegraphics[width=\columnwidth]{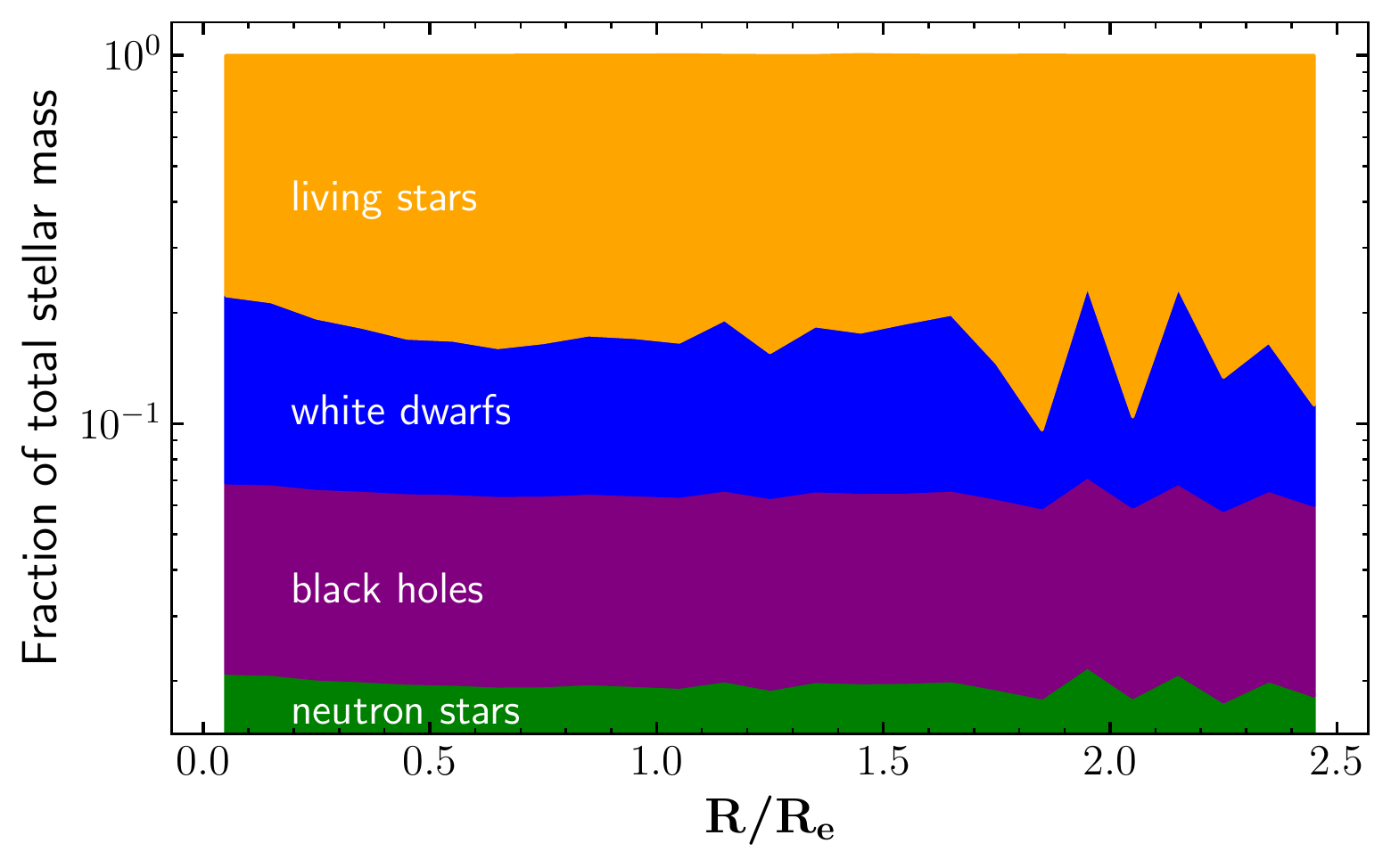}
    		\caption{Average fractional contribution of stellar remnants to the total stellar mass per spatial bin as a function of radius. Due to the large parameter range, we show fractions in log units. Based on the 300 galaxy subsample introduced in Sect. \ref{sect:config_tests}.}
    		\label{fig:remnants}
		\end{figure}
		
		\begin{figure}
			\centering
			\includegraphics[width=\columnwidth]{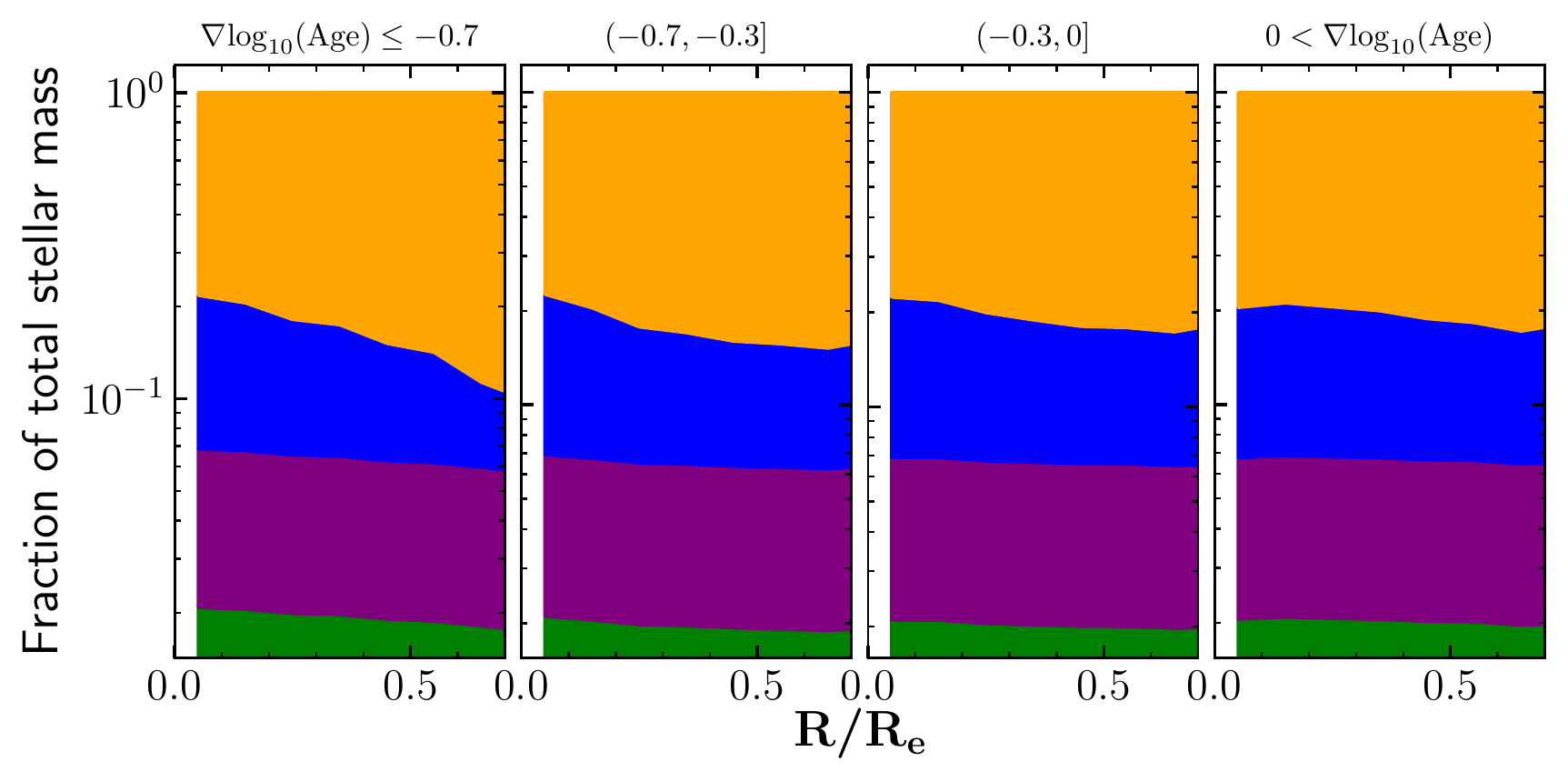}
    		\caption{Same as Fig. \ref{fig:remnants} but for galaxies divided into four different bins of radial stellar age gradient. Shown are only the inner $0.7\,R_\mathrm{e}$.}
    		\label{fig:remnants2}
		\end{figure}
		
		In addition to stellar ages, metallicities, masses, dust attenuation and star formation histories, the \texttt{MaNGA {\sc firefly} VAC} also contains spatially resolved information of stellar remnants. These include black holes, neutron stars and white dwarfs. The masses of the remnants are derived by applying mass loss factors to the SSPs \citep{Maraston2005}, as described in Sect. \ref{sect:ff_run}. An example showing 2D maps of the total mass of stellar remnants is shown in Fig. \ref{fig:maps}.
		
		Another instructive way to look at these quantities is shown in Fig. \ref{fig:remnants}. This plot shows the fraction of stellar remnants to the total stellar mass as a function of radius. All fractions are calculated as the average of all spatial bins of the 300 galaxy test sample from the previous subsection. Interestingly, the fractional mass of all remnants is slightly radially decreasing between the centre and $R=0.7\,R_\mathrm{e}$, most prominently seen for white dwarfs. The reason for that might be related to radially decreasing age, as younger populations contain more living stars. To test this hypothesis, we bin the galaxies in four groups of different radial light-weighted stellar age gradient intervals and show the same plot for each group of galaxies in Fig. \ref{fig:remnants2}. Indeed, galaxies with the steepest age gradient, have also the strongest radial decrease of remnants. Figures \ref{fig:remnants} and \ref{fig:remnants2} exemplify the potential of the data contained in the VAC. Detailed maps of stellar remnants will be presented and discussed in a forthcoming paper.

\section{Performance}
\label{sect:perf}

		\begin{figure*}
			\centering
			\includegraphics[width=17cm]{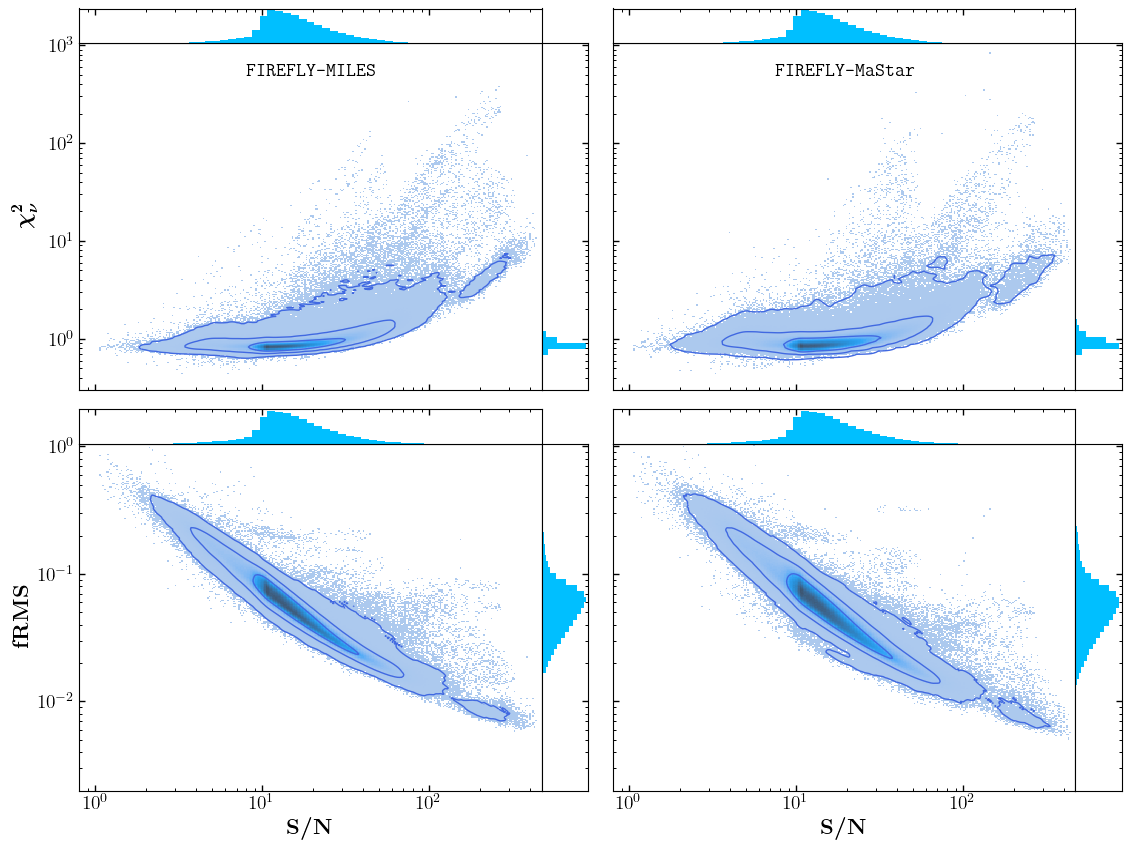}
    		\caption{Goodness of FIREFLY fits to MaNGA data. Shown are the reduced $\chi_\nu^2$ as well as the root mean square of the fractional residuals $f\mathrm{RMS}$ as a function of $g$-band S/N. The complete sample of spectral fits is included. 2D densities and marginals are in linear scales with darker regions corresponding to higher densities. Contours enclose 1$\sigma$, 2$\sigma$ and 3$\sigma$ of the distribution.}
    		\label{fig:perf}
		\end{figure*}
		
	The \texttt{MaNGA {\sc firefly} VAC} does not explicitly provide any quality flags of the data products. On the one hand, the input data have already been subject to quality control in the processing of the {\sc drp} and {\sc dap}. Only successful {\sc dap} output was considered for further {\sc firefly} fitting. We strongly recommend to use the data products in combination with quality control flags as provided by the {\sc drp} and {\sc dap}. On the other hand, all output parameters in the \texttt{VAC} are paired with an error estimation as detailed in Sect. \ref{sect:workflow}. In addition, in this section, we study the general quality of the fits by considering the individual spectral residuals between observations and best-fitting models. These measurements are not part of the official \texttt{VAC} release but will be made publicly available alongside the \texttt{VAC} at \url{http://www.icg.port.ac.uk/manga-firefly-vac}.
		
	The procedure to characterise the goodness of the spectral fits in the \texttt{MaNGA {\sc firefly} VAC} is for the most part adapted from \citet{Westfall2019}. We calculate for each fit the fractional residual per wavelength channel $\Delta_{f,i} = |f_{\mathrm{obs},i}-f_{\mathrm{model},i}|/f_{\mathrm{model},i}$ and the error-normalised residual $\Delta_{\mathcal{E},i} = |f_{obs,i}-f_{\mathrm{model},i}|/\mathcal{E}_{i}$. Here, $f_i$ is the flux of the observed or best-fit model spectrum at wavelength channel $i$, and $\mathcal{E}_{i}$ is the corresponding error of the observed flux. We define then the fractional root mean square $f\mathrm{RMS}$ and the reduced $\chi_\nu^2$ as
	
	\begin{align}
	f\mathrm{RMS} &= \sqrt{\frac{1}{N} \sum_{i=1}^N {\Delta_{f,i}^2}} ,\\
	\chi_\nu^2 &= \left. \sum_{i=1}^N {\Delta_{\mathcal{E},i}^2}  \middle/ (N-\nu) \right. ,
	\end{align}
	
	where N is the total number of fitted wavelength channels and $\nu$ is the number of model templates with non-zero weight. In Fig. \ref{fig:perf}, we plot $f\mathrm{RMS}$ and $\chi_\nu^2$ against the $g$-band S/N per fit, in other words per Voronoi bin. For a theoretically perfect fit, we expect an anti-correlation between $f\mathrm{RMS}$ and S/N and a constant, close-to-one $\chi_\nu^2$.
	
	The observed trends in Fig. \ref{fig:perf} are very similar to the ones seen for the stellar continuum fits by the {\sc dap} and are discussed in detail in \citet[][their figure 27 and section 11.2.1]{Westfall2019}. The $f\mathrm{RMS}$ follow a clear log-log anti-correlation with S/N. We observe a slight flattening in the relation at high S/N, which is paralleled by an up-bending of $\chi_\nu^2$. At these S/N systematic errors in the modelling start to dominate random errors. Overall, the distribution of $\chi_\nu^2$ is mostly flat with a median value of $\langle \chi_\nu^2\rangle = 0.87\pm0.05$ for \texttt{FF-Mi} and $\langle \chi_\nu^2\rangle = 0.90\pm0.07$ for \texttt{FF-Ma}. The fact that $\chi_\nu^2$ is on average smaller than one could be a sign of overfitting, but is rather likely due to an overestimation of the flux error \citep[c.f.][]{Westfall2019}. There is no significant difference in the performance of the fits between \texttt{FF-Mi} and \texttt{FF-Ma}. The median $\chi_\nu^2$ value of \texttt{FF-Ma} is higher by 0.03, which is probably caused by the noisier part of the spectra at large wavelength covered by the MaStar models. In fact, when the MaStar fits are limited to the MILES wavelength range (\texttt{config5}), the median $\chi_\nu^2$ drops to $0.85\pm0.05$.
		
\section{Comparison with other MaNGA catalogues}
\label{sect:literature}

		\begin{figure*}
			\centering
			\includegraphics[width=17cm]{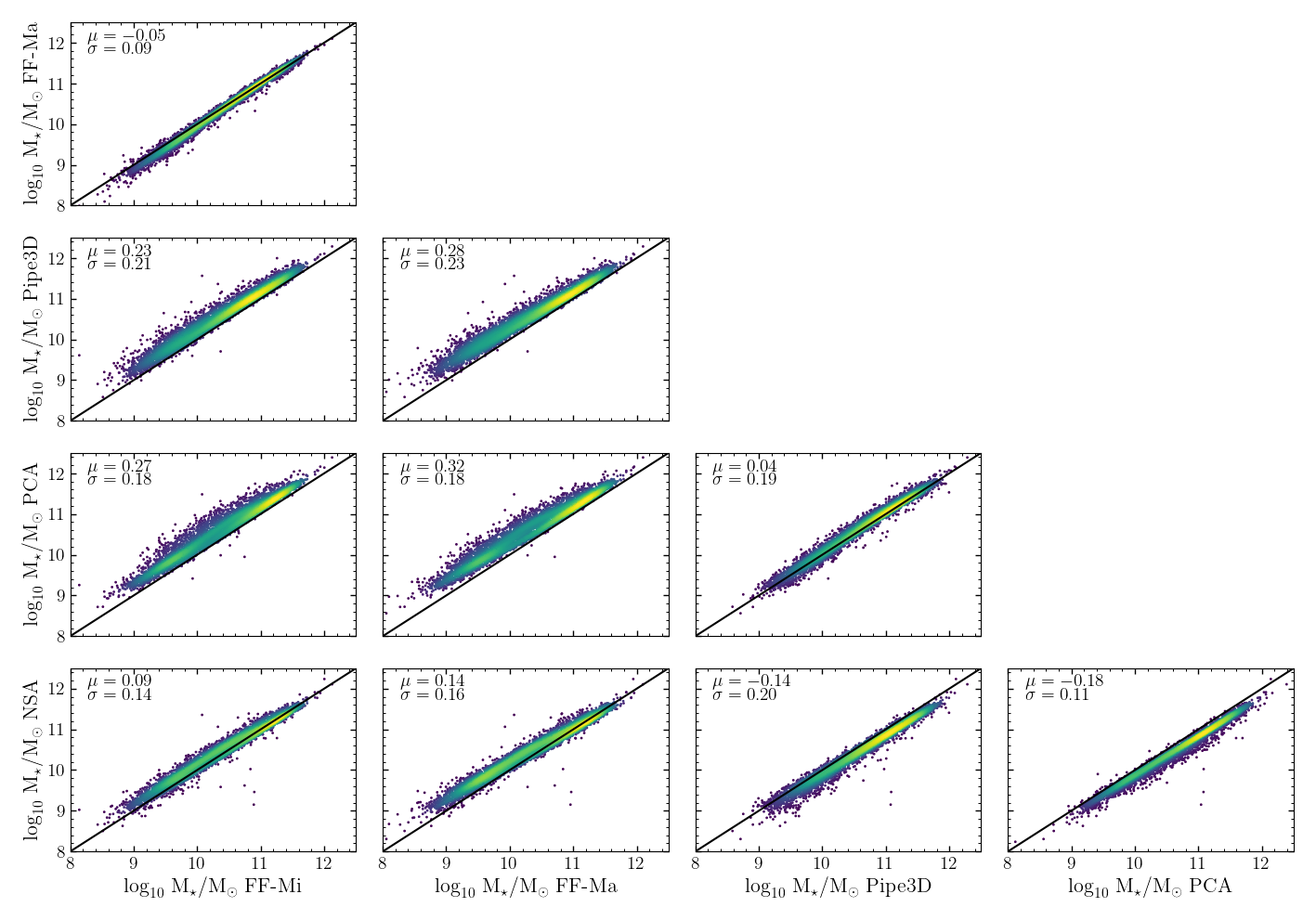}
    		\caption{Total stellar masses of DR17 MaNGA galaxies as listed in different catalogues. All masses are converted to a cosmology with $H_0=67.8\,\mathrm{km\,s^{-1}\,Mpc^{-1}}$, Kroupa IMF and are aperture corrected.}
    		\label{fig:mass}
		\end{figure*}

A total number of 63 Value-Added-Catalogues have been released in SDSS as of DR17, of which 25 are updated or new in DR17 \citep{Abdurrouf2021}. This includes 3 stellar population modelling VACs of MaNGA data: the Principle Component Analysis VAC \citep[\texttt{PCA};][]{Pace2019a,Pace2019b}, the \texttt{Pipe3D} VAC (S\'anchez et al., in prep., see \citealp{Sanchez2018} for earlier versions) based on the new pipeline version \texttt{pyPipe3D} \citep{Lacerda2022} originally described in \citet{Sanchez2016a}, and the \texttt{MaNGA {\sc firefly} VAC}. In this section, we compare some of our results with the DR17 version of these VACs and output from the {\sc dap}.

	\subsection{Total galaxy masses}
	\label{sect:masses}

	The mass is a fundamental parameter in the characterisation of a galaxy and is one of the main drivers of physical processes as part of both the dynamical and chemical evolution \citep[e.g.][]{Kauffmann2003,Thomas2010,Peng2010}. The determination of the stellar mass through full spectral fitting with stellar population models is usually more robust to the exact combination of SSP templates than other parameters such as age or metallicity because of degeneracy effects. While there is indeed little scatter in mass between the configuration tests performed in the previous section, the absolute calibration of stellar masses depend on several assumptions, e.g. the cosmological parameters, the IMF and the input physics used in the population synthesis code.
	
	Spatially resolved stellar masses in the MaNGA sample are available in the \texttt{MaNGA {\sc firefly} VAC} variants (1) \texttt{FF-Mi} and (2) \texttt{FF-Ma}, (3) in the \texttt{PCA} VAC, (4) in the \texttt{Pipe3D} VAC, and (5) total galaxy masses in the NASA SLoan Atlas \citep[NSA;][]{Blanton2011,Wake2017}. For future reference when comparing MaNGA works based on stellar mass measurements from different sources, we show in Fig. \ref{fig:mass} a cross-comparison between each of the catalogues.
	
	A number of notes have to be made. The MaNGA VACs (1)-(4) are based on resolved IFU data, but use mutually different binning schemes. Therefore, we decided to only compare \textit{total} galaxy masses that are additionally also comparable to the photometry-based measurement from the NSA catalogue. The total mass in (1)-(4) is simply the sum across the MaNGA field-of-view. All masses are converted to the cosmology used in the {\sc firefly} VAC, i.e. \citet{Planck2016}. Furthermore we shift all \texttt{Pipe3D} masses, which assume a Salpeter IMF, by 0.17\,dex using the offset found in Sect. \ref{sect:imf} so that all catalogues use Kroupa IMF equivalent masses. The sample in (1), (2), (4) and (5) is the DR17 sample. The comparison to (3) is based on the common DR15 subsample, since the DR17 version of the \texttt{PCA} VAC was not yet available at the writing of this paper. Finally, we apply an aperture correction to all IFU-based measurements, i.e. VACs (1)-(4), using the correction factors based on the \textit{CMLR} method from \citet{Pace2019b}. In this method the mass of each galaxy outside the IFU is calculated from the missing flux using colour-mass-to-light relations \citep{Pace2019a}.
	
	The scatter between any pair of catalogues in Fig. \ref{fig:mass} is on the order of 0.2\,dex with the exception of the two {\sc firefly} VACs with $\sigma=0.09$\,dex. There are clear systematic offsets between all catalogues: As compared to \texttt{FF-Ma} with the lowest masses, the offsets in dex are 0.05 for \texttt{FF-Mi}, 0.14 for \texttt{NSA}, 0.28 for \texttt{Pipe3D} and 0.32 for \texttt{PCA}, which reports the highest masses.  The stellar masses published in the \texttt{Pipe3D} VAC are masses of living stars only. The offset reported here is therefore likely to increase by $\sim 0.1$\,dex (c.f. Fig. \ref{fig:remnants}) if remnants are included.
	
	Differences between these catalogues are likely due to different input libraries, stellar tracks and mass loss prescriptions. The \texttt{PCA} VAC uses an unpublished theoretical stellar library and Padova 2008 \citep{Marigo2008} isochrones to synthesise the models used in the VAC. The DR17 version of the \texttt{Pipe3D} VAC uses an unpublished model library based on the MaStar stellar library using a code from Bruzual \& Charlot (BC19, priv. comm.). Hence, at this point it is difficult to pin-point the exact sources leading to the discrepancies, but the different inputs in the model libraries are the most likely reason.

	\subsection{Stellar ages and metallicities from Pipe3D}
	
		\begin{figure}
			\centering
			\includegraphics[width=\columnwidth]{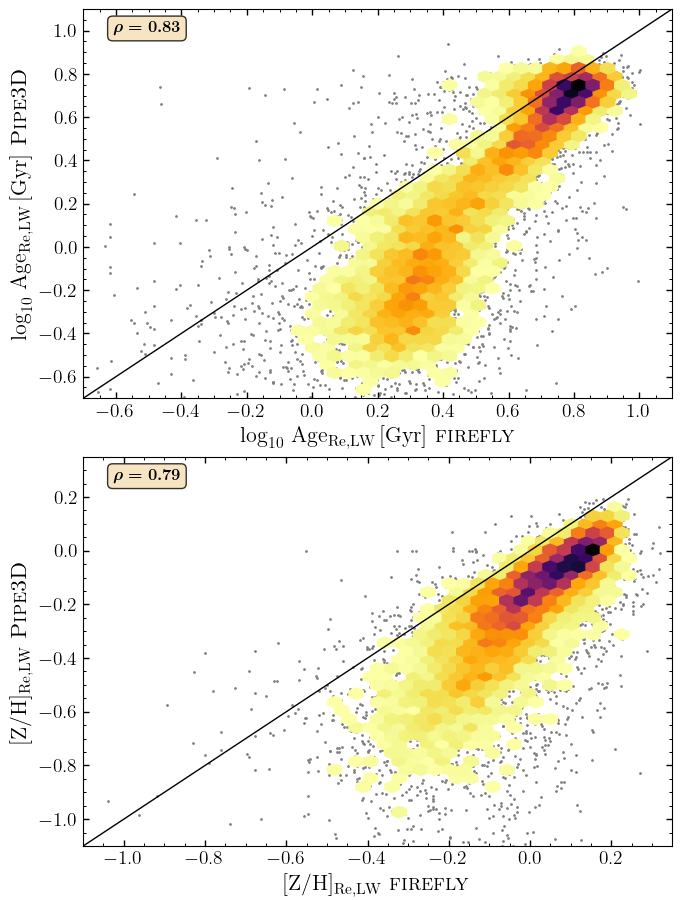}
    		\caption{Comparison of light-weighted stellar ages and metallicities averaged within a ring at $1\,R_\mathrm{e}$ between \texttt{Pipe3D} and {\sc firefly} (\texttt{FF-Ma}).}
    		\label{fig:pipe3d}
		\end{figure}

		The \texttt{Pipe3D} VAC comprises an extensive catalogue of stellar populations and emission line properties of MaNGA galaxies that are derived independently and in parallel to the properties of the {\sc dap} and the {\sc firefly} VAC. An in depth comparison between both VACs is very desirable and necessary. Nevertheless, there are a few complications to overcome. Firstly, both VACs use different Voronoi binning schemes. Therefore, fitting is performed on non-identical spectra and a direct one-to-one comparison of the stellar population properties is not possible. Instead, we will focus on averaged `global' galaxy properties. Secondly, the comparison of star formation histories is a complex multi-dimensional problem and, therefore, comparisons must rely on averaged stellar populations. \texttt{Pipe3D} uses geometric means while the {\sc firefly} \texttt{VAC} uses arithmetic means. The geometric mean is always lower than the arithmetic mean by an amount that correlates with the variance in the data. It is beyond the scope of this paper to recalculate the means from the star formation histories. Nonetheless, it is still instructive to look for correlation between both VACs in a qualitative way.
		
		In Fig. \ref{fig:pipe3d}, we compare the mean light-weighted age and metallicity at the effective radius for the complete sample of $\sim 10,000$ galaxies between \texttt{FF-Ma} and \texttt{Pipe3D}, of which both VACs are based on the MaStar stellar library. The \texttt{Pipe3D} parameter is averaged between two concentric ellipses at 0.75 and $1.25\,R_\mathrm{e}$ while in the \texttt{MaNGA {\sc firefly} VAC}, we average between 0.9 and $1.1\,R_\mathrm{e}$. As expected, \texttt{Pipe3D}-derived mean ages and metallicities are systematically lower, most probably because of the usage of geometric means. However, we reiterate and note that both VACs adopt different stellar population models. The systematic difference in Fig. \ref{fig:pipe3d} is larger at lower values, which is likely caused by a larger spread in the star formation histories.  While a comparison of the total values is therefore impeded, we find a strong monotonic correlation as attested by the Spearman's rank correlation coefficients of $\rho=0.83$ and $\rho=0.79$. We acknowledge this as encouraging indication of a good agreement.\footnote{The results of both stellar population VACs are not expected to fully agree given the different approaches in the astrophysical modelling involved in the procedure. Nonetheless, it is essential to explore discrepancies and congruencies.} Further and more detailed studies are needed to fully explore both VACs in comparison.

	\subsection{Star formation rates from Pipe3D}
	\label{sect:sfr_pipe3d}
	
	Local and global star formation activity is one of the main parameters used to characterise galaxies on large scales and to learn about star formation processes on sub-kpc scales. MaNGA provides the opportunity to measure both spatially resolved and global star formation of 10,010 galaxies. A variety of star formation tracers across the electromagnetic spectrum are frequently been used, most notably H$\alpha$ as the strongest recombination line tracing the ionisation of young, massive stars.
	
	A full stellar population analysis as performed in this VAC offers probably the most direct probe of recent star formation by delivering a full decomposition of stellar ages. The SFR in the \texttt{MaNGA {\sc firefly} VAC} is calculated as sum over all mass of stellar populations younger than $10\,\mathrm{Myr}$ and $100\,\mathrm{Myr}$ divided by the corresponding time interval (see Sect. \ref{sect:sfr}). Similarly, \texttt{Pipe3D} provides SFR measurements averaged over $10\,\mathrm{Myr}$, $32\,\mathrm{Myr}$ and $100\,\mathrm{Myr}$. In addition, \texttt{Pipe3D} also provides the SFR based on dust-corrected H$\alpha$ flux measurements. As we did for the stellar age and metallicity comparison, we compare our SFRs to the \texttt{Pipe3D}  H$\alpha$-based SFRs on a global, per-galaxy basis, since individual binning schemes differ between both VACs.
	
\begin{figure*}
			\centering
			\includegraphics[width=17cm]{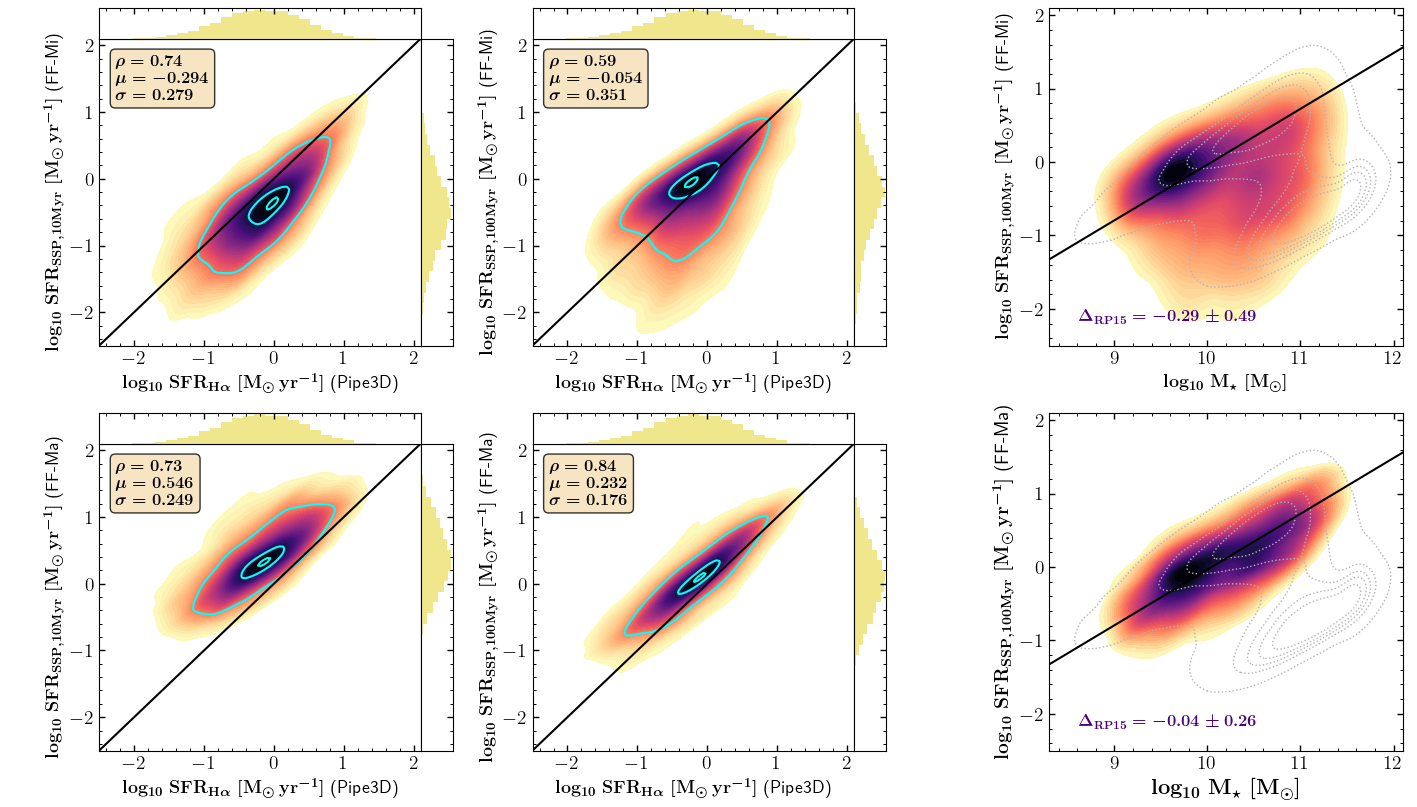}
    		\caption{\textit{Left and middle column:} Comparison of SFRs between the SSP-based approach in the \texttt{MaNGA {\sc firefly} VAC} and the H$\alpha$-based SFRs from \texttt{Pipe3D}. Shown are SFRs considering a $10\,\mathrm{Myr}$ and a $100\,\mathrm{Myr}$ time interval. Colours, contours, histograms and statistical parameters are as in Fig. \ref{fig:mi_vs_ma}. \textit{Right column:} SFR-M$_\star$ plot. Colours show the density distribution of our SSP-based measurements, while grey contours indicate the distribution of SDSS galaxies \citep{Brinchmann2004}. The black solid line marks the star formation main sequence as determined in \citet{Renzini2015}. Our sample is limited to galaxies with $\mathrm{|EW(H\alpha)|}>3\,\text{\AA}$ at 1 $R_\mathrm{e}$.}
    		\label{fig:sfr}
		\end{figure*}

	For this comparison, we only select star-forming galaxies with clear H$\alpha$ detection, characterised by $\mathrm{|EW(H\alpha)|}>3\,\text{\AA}$ at 1 $R_\mathrm{e}$, following S\'anchez et al. (in prep.). Figure \ref{fig:sfr} shows that our SSP-based SFRs generally correlate well with H$\alpha$-based SFRs from \texttt{Pipe3D}. The best correlation with the lowest scatter is found for $\mathrm{SFR_{SSP,100\,Myr}}$ in \texttt{FF-Ma} with an offset of $\Delta \log_{10}\mathrm{SFR_{SSP-H\alpha}} = 0.232\pm0.176$. \texttt{FF-Mi} likely underestimates SFRs due to the lack of young SSP templates. The effect is larger for the younger (shorter) $10\,\mathrm{Myr}$ range. Despite the offset of \texttt{FF-Ma} SFRs, we find that their distribution agrees exceptionally well with the main sequence of star formation as shown in the right-hand panels, where we overplot our star-forming sample with the contours from the distribution of DR7 SDSS galaxies based on \citet{Brinchmann2004}. The median difference to the main sequence as parametrised in \citet{Renzini2015} is $\Delta_\mathrm{RP15} = -0.04\pm0.26$.
	
	\subsection{Spectral indices and dust attenuation from the {\sc dap}}
	\label{sect:dap}

		\begin{figure*}
			\centering
			\includegraphics[width=18cm]{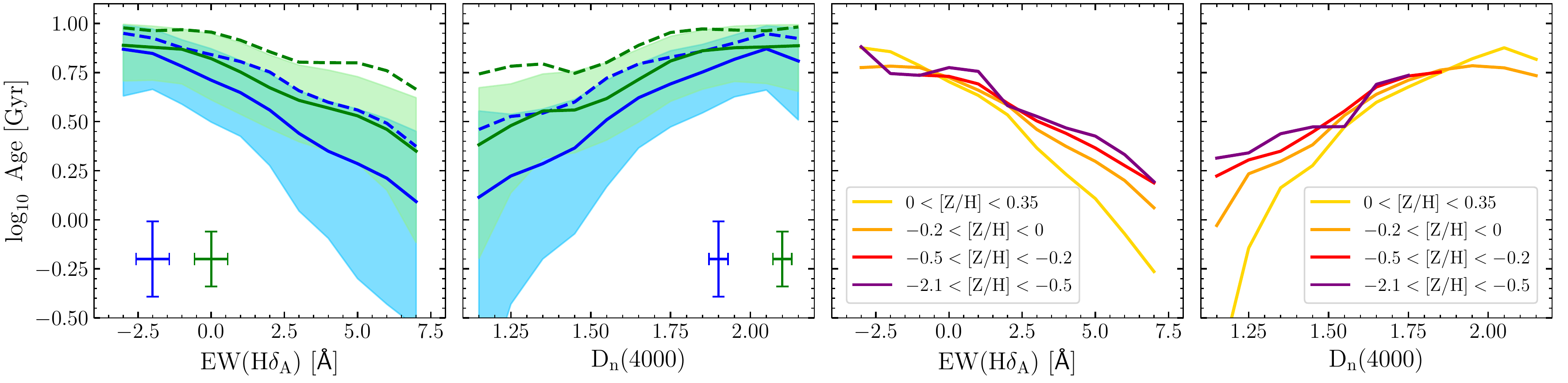}
    		\caption{Comparison between {\sc firefly} derived stellar ages and metallicities with spectral indices from the {\sc dap}. \textit{Left two panels:} Stellar ages versus EW(H$\delta$) and $\rm D_n(4000)$. Values from \texttt{FF-Ma} are shown in blue, \texttt{FF-Mi} is shown in green. Solid lines are light-weighted, dashed lines are mass-weighted median ages. Regions between the 16th and 84th percentile of light-weighted values are shown as shades. Errorbars show the median error of individual measurements. \textit{Right two panels:} Same as left, but only showing light-weighted ages separated in bins of stellar metallicity. This figure shows the results from the subsample of 300 galaxies introduced in Sect. \ref{sect:config_tests}.}
    		\label{fig:dap}
		\end{figure*}
	
		\begin{figure}
			\centering
			\includegraphics[width=\columnwidth]{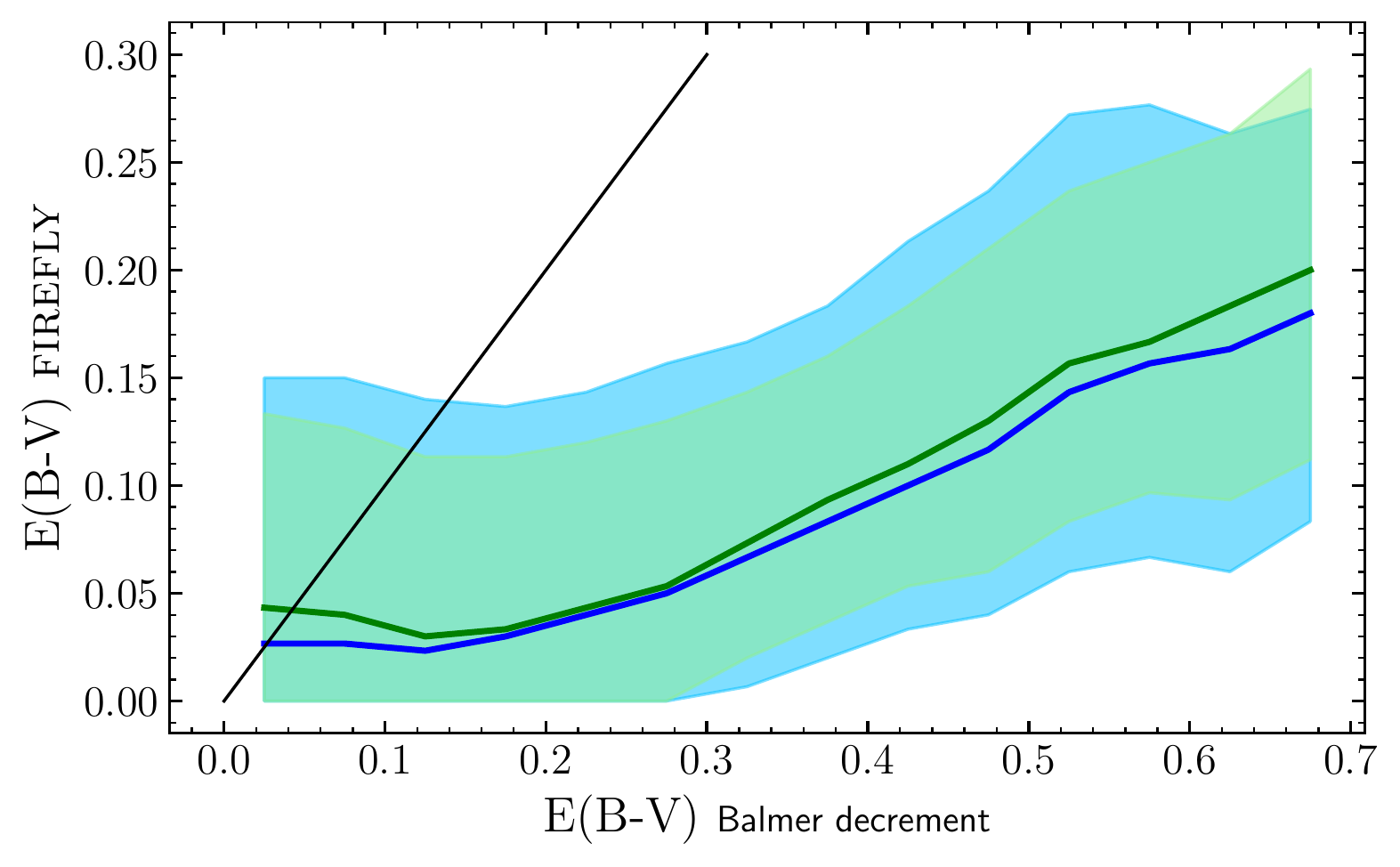}
    		\caption{Comparison of colour excess measured from stellar population fitting ({\sc firefly}) and emission line analysis ({\sc dap}). The black line shows the one-to-one relation.}
    		\label{fig:dap_ebv}
		\end{figure}
	
		The \texttt{MaNGA {\sc firefly} VAC} is built upon the {\sc dap}. One of the advantages of that is that both provide data products for identical spectra from identical spatial bins. For that reason, spatially resolved parameters can be directly compared to each other.
	
		The following figures can be understood as kind of sanity checks and as explorations of the general distributions of parameters in the catalogues rather than deep scientific analyses, which would require more careful sample selections, adjustments and appropriate in-depths discussions.
		
		In the first two columns of Fig. \ref{fig:dap}, we plot {\sc firefly} derived stellar ages against two commonly used age indicators \citep{Poggianti1997,Worthey1997,Bruzual2003,Kauffmann2003} as measured by the {\sc dap}: the equivalent width of the Balmer absorption line H$\delta$ and the $4000\,\text{\AA}$ break index $\rm D_n(4000)$ \citep{Balogh1999}. This analysis is again based on single spatial bins of the 300 galaxy subsample described previously. We see that age anti-correlates with H$\delta$ and correlates with $\rm D_n(4000)$. The correlation with light-weighted ages is steeper and stronger than with mass-weighted ages (Spearman's rank $\rho=0.70$ and $\rho=0.55$ for H$\delta$; $\rho=0.68$ and $\rho=0.53$ for $\rm D_n(4000)$). This confirms expectations, because both indices are most sensible to stellar ages below 2 Gyr \citep[e.g.][]{Bruzual2003} and changes in the young stellar populations dominate light-weighted ages more strongly. \citet{Maraston2005} pointed out that $\rm D_n(4000)$ is not a pure age indicator but also sensible to changes in metallicity. To test this, we plot the same relation between age and both indices binned in four metallicity ranges. In agreement with \citet{Maraston2005}, at a given age, different H$\delta$ and $\rm D_n(4000)$ are found for varying metallicity. The most metal-poor populations correspond to the lowest values of H$\delta$ and the highest values of $\rm D_n(4000)$. Care should be taken when using them as pure age indicator.
		
		Figure \ref{fig:dap_ebv} presents a comparison between the colour excess $E(B-V)$, on the one hand, derived as part of the full spectral stellar population fitting process by {\sc firefly} and, on the other hand, derived from the emission line measurements in the {\sc dap} using the reddening of the Balmer decrement and assuming a Case B recombination $E(B-V)=1.97 \log_{10}\, [(\mathrm{H\alpha/H\beta})/2.86]$ \citep{Baker1938,Osterbrock1989}. Despite the fact that both colour excesses are measured in two completely different and independent ways, we find a clear positive correlation, yet with a large amount of scatter. Furthermore, the {\sc firefly} values are clearly below the one-to-one correlation for both variants of the VAC. The reddening from the Balmer decrement is expected to be higher because it captures the close, dusty environment of young stellar populations.

\section{Summary}
\label{sect:summary}

We have presented the DR17 \texttt{MaNGA {\sc firefly} Value-Added-Catalogue (VAC)} in its two variants \texttt{FF-Mi} and \texttt{FF-Ma}; a catalogue of spatially resolved stellar population properties of $10,010$ nearby galaxies from the MaNGA survey, obtained through state-of-the-art full spectral fitting of data with stellar population models. Both variants are structurally identical catalogues of the same sample, once fitted with \texttt{M11-MILES} and once with the \texttt{MaStar} stellar population models, respectively. Parameters provided in the \texttt{VAC} include stellar ages, metallicities, stellar and remnant masses, star formation histories and dust attenuation; a full list of the content can be found in Table \ref{tbl:vac_content}. The \texttt{VAC} is a major update to earlier data release versions \citep{Goddard2017} and this paper is the first complete description of the \texttt{VAC}. It contains updates to the {\sc firefly} \citep{Wilkinson2017} code itself as well as further elements of the parameter estimation and \texttt{VAC} construction.

A schematic illustration of the workflow can be found in Fig. \ref{fig:workflow}, example fits in Figs. \ref{fig:spec1} and \ref{fig:spec2} and a comparison of the distribution of stellar population properties in Fig. \ref{fig:mi_vs_ma}.

The major new additions of this \texttt{VAC} is the inclusion of the \texttt{FF-Ma} variant employing the novel \texttt{MaStar} models (\citealp{Maraston2020}, Maraston et al. 2022, in prep.) based on the MaStar stellar library \citep{Yan2019} observed with the same instrument over the same wavelength range as the MaNGA data. The same \texttt{MaStar} models are also used as input for the data analysis pipleline \citep{Westfall2019}. Owing to the large sample size in MaStar the parameter space is much better sampled allowing for an extended model grid including low-age low-metallicity templates.

We perform a variety of fitting tests and comparisons with different configurations in order to aid understanding the fitted stellar population parameters included in this \texttt{VAC}. \texttt{FF-Ma} provides on average slightly younger ages, higher mass-weighted metallicities and smaller colour excesses than \texttt{FF-Mi}. These differences are reduced when matching the wavelengths range and model parameter grid. We further provide a comparison between the {\sc firefly} \texttt{VAC} and other MaNGA stellar population catalogues. The masses in the \texttt{MaNGA {\sc firefly} VAC} are systematically lower by $\sim 0.3\,\mathrm{dex}$, but match the photometrically derived masses in the \texttt{NSA} catalogue best.

This version of the \texttt{VAC} has been published as an official \texttt{VAC} together with DR17 of SDSS. Performance tests (Sect. \ref{sect:perf}) and the calculation of star formation rates (Sect. \ref{sect:sfr} and \ref{sect:sfr_pipe3d}) have been performed after the official release. Furthermore, we foresee to update the catalogue with new variants employing later versions of the \texttt{MaStar} model library and we are also looking into using variable IMFs as a free parameter in the fitting procedure in future versions. SFRs and updates to the DR17 \texttt{VAC} will be published on the ICG Portsmouth institute's website: \url{http://www.icg.port.ac.uk/manga-firefly-vac}.

\section*{Acknowledgements}

We thank the anonymous referee for a very constructive report that helped to improve the paper. JN thanks his wife Karina for the design of the workflow illustration. The Science and Technology Facilities Council is acknowledged for support through the Consolidated Grant Cosmology and Astrophysics at Portsmouth, ST/S000550/1. VGP is supported by the Atracci\'{o}n de Talento Contract no. 2019-T1/TIC-12702 granted by the Comunidad de Madrid in Spain. RY acknowledges financial support by the Hong Kong Global STEM Scholar scheme and the Direct Grant of CUHK Faculty of Science.

Numerical computations were done on the Sciama High Performance Compute (HPC) cluster which is supported by the ICG, SEPnet and the University of Portsmouth.

Funding for the Sloan Digital Sky Survey IV has been provided by the Alfred P. Sloan Foundation, the U.S. Department of Energy Office of Science, and the Participating Institutions. SDSS acknowledges support and resources from the Center for High-Performance Computing at the University of Utah. The SDSS web site is \url{www.sdss.org}.

SDSS is managed by the Astrophysical Research Consortium for the Participating Institutions of the SDSS Collaboration including the Brazilian Participation Group, the Carnegie Institution for Science, Carnegie Mellon University, the Chilean Participation Group, the French Participation Group, Harvard-Smithsonian Centre for Astrophysics, Instituto de Astrof\'isica de Canarias, The Johns Hopkins University, Kavli Institute for the Physics and Mathematics of the Universe (IPMU) / University of Tokyo, the Korean Participation Group, Lawrence Berkeley National Laboratory, Leibniz Institut f\"ur Astrophysik Potsdam (AIP), Max-Planck-Institut f\"ur Astronomie (MPIA Heidelberg), Max-Planck-Institut f\"ur Astrophysik (MPA Garching), Max-Planck-Institut f\"ur Extraterrestrische Physik (MPE), National Astronomical Observatories of China, New Mexico State University, New York University, University of Notre Dame, Observatório Nacional / MCTI, The Ohio State University, Pennsylvania State University, Shanghai Astronomical Observatory, United Kingdom Participation Group, Universidad Nacional Aut\'onoma de M\'exico, University of Arizona, University of Colorado Boulder, University of Oxford, University of Portsmouth, University of Utah, University of Virginia, University of Washington, University of Wisconsin, Vanderbilt University, and Yale University.

\section*{Data availability}

The \texttt{MaNGA {\sc firefly} VAC} described in this paper can be downloaded from the SDSS website \url{https://www.sdss.org/dr17/manga/manga-data/manga-firefly-value-added-catalog} or from the ICG Portsmouth's website \url{http://www.icg.port.ac.uk/manga-firefly-vac}. It is also available as {\sc cas} table on the SDSS skyserver \url{http://skyserver.sdss.org/dr17} and integrated in {\sc marvin} \url{https://dr17.sdss.org/marvin}. The {\sc firefly} code is publicly available at \url{https://github.com/FireflySpectra/firefly_release} and is described at \url{https://www.icg.port.ac.uk/firefly} and \url{https://www.sdss.org/dr17/spectro/galaxy_firefly}. The stellar population models used in this paper are available at \url{https://svn.sdss.org/public/data/sdss/stellarpopmodels/tags/v1_0_2/} and \url{http://www.icg.port.ac.uk/mastar} and also integrated in the {\sc firefly} github package.

%%%%%%%%%%%%%%%%%%%%%%%%%%%%%%%%%%%%%%%%%%%%%%%%%%



\bibliographystyle{mnras}
\bibliography{../../NewDatabase.bib} % if your bibtex file is called example.bib

%%%%%%%%%%%%%%%%%%%%%%%%%%%%%%%%%%%%%%%%%%%%%%%%%%

%%%%%%%%%%%%%%%%% APPENDICES %%%%%%%%%%%%%%%%%%%%%

\appendix

\section{VAC Content}
\label{sect:vac_content}

Here we present the content of the \texttt{MaNGA {\sc firefly} VAC} in Table \ref{tbl:vac_content}.

\begin{table*}
\caption{MaNGA {\sc firefly} VAC - Content of Catalogue}
\label{tbl:vac_content}
\centering

\begin{tabular}{l p{0.2\linewidth} l p{0.06\linewidth} >{\hangindent=2em}p{0.52\linewidth}}
\hline\hline
HDU & Name & Dimensions & Units & Description\\
(1) & (2) & (3) & (4) & (5)\\
\hline
0 & PRIMARY & & & Empty primary header.\\
\hline
1 & MANGAID & & & Unique MaNGA identifier.\\
& PLATEIFU & & & Unique identifier containing the MaNGA plate and ifu combination.\\
& PLATE & & & Plate used to observe galaxy.\\
& IFUDSGN & & & IFU used to observe galaxy.\\
& OBJRA & & & Right ascension of the galaxy, not the IFU.\\
& OBJDEC & & & Declination of the galaxy, not the IFU.\\
& REDSHIFT & & & Redshift of the galaxy.\\
& PHOTOMETRIC\_MASS & & $\rm \log_{10}\,(M_\odot)$ & Stellar mass of galaxy from NSA catalogue obtained from K-correction fits to elliptical Petrosian photometric fluxes. Masses are converted to the cosmology parameters that we assume in the {\sc firefly} \texttt{VAC} with $H_0 = 67.8\,\mathrm{km\,s^{-1}\,Mpc^{-1}}$.\\
& MANGADRP\_VER & & & Version of MaNGA DRP that produced this data.\\
& MANGADAP\_VER & & & Version of MaNGA DAP that analysed this data.\\
& FIREFLY\_VER & & & Version of FIREFLY that analysed this data.\\
\hline

2 & LW\_AGE\_1Re & & $\rm \log_{10}\,(M_\odot)$ & Light-weighted age within a shell located at $1\,\mathrm{R_e}$.\\
 & LW\_AGE\_1Re\_ERROR & & $\rm \log_{10}\,(M_\odot)$ & Error on light-weighted age within a shell located at $1\,\mathrm{R_e}$.\\
 & MW\_AGE\_1Re & & $\rm \log_{10}\,(M_\odot)$ & Mass-weighted age within a shell located at $1\,\mathrm{R_e}$.\\
 & MW\_AGE\_1Re\_ERROR & & $\rm \log_{10}\,(M_\odot)$ & Error on mass-weighted age within a shell located at $1\,\mathrm{R_e}$.\\
 & LW\_Z\_1Re & & & Light-weighted metallicity [Z/H] within a shell located at $1\,\mathrm{R_e}$.\\
 & LW\_Z\_1Re\_ERROR & & & Error on light-weighted metallicity [Z/H] within a shell located at $1\,\mathrm{R_e}$.\\
 & MW\_Z\_1Re & & & Mass-weighted metallicity [Z/H] within a shell located at $1\,\mathrm{R_e}$.\\
 & MW\_Z\_1Re\_ERROR & & & Error on mass-weighted metallicity [Z/H] within a shell located at $1\,\mathrm{R_e}$.\\
 & LW\_AGE\_3ARCSEC & & $\rm \log_{10}\,(M_\odot)$ & Light-weighted age within $3\arcsec$ diameter.\\
 & LW\_AGE\_3ARCSEC\_ERROR & & $\rm \log_{10}\,(M_\odot)$ & Error on light-weighted age within $3\arcsec$.\\
 & MW\_AGE\_3ARCSEC & & $\rm \log_{10}\,(M_\odot)$ & Mass-weighted age within $3\arcsec$ diameter.\\
 & MW\_AGE\_3ARCSEC\_ERROR & & $\rm \log_{10}\,(M_\odot)$ & Error on mass-weighted age within $3\arcsec$.\\
 & LW\_Z\_3ARCSEC & & & Light-weighted metallicity [Z/H] within $3\arcsec$ diameter.\\
 & LW\_Z\_3ARCSEC\_ERROR & & & Error on light-weighted metallicity [Z/H] within $3\arcsec$.\\
 & MW\_Z\_3ARCSEC & & & Mass-weighted metallicity [Z/H] within $3\arcsec$ diameter.\\
 & MW\_Z\_3ARCSEC\_ERROR & & & Error on mass-weighted metallicity [Z/H] within $3\arcsec$.\\
\hline

3 & LW\_AGE\_GRADIENT & & dex/$\mathrm{R_e}$	& Light-weighted age gradient of linear fit obtained within $1.5\,\mathrm{R_e}$.\\
 & LW\_AGE\_GRADIENT\_ERROR & & dex/$\mathrm{R_e}$	& Error on light-weighted age gradient within $1.5\,\mathrm{R_e}$.\\
 & LW\_AGE\_ZEROPOINT & & & Light-weighted age zeropoint of linear fit obtained within $1.5\,\mathrm{R_e}$.\\
 & LW\_AGE\_ZEROPOINT\_ERROR & & & Error on light-weighted age zeropoint obtained within $1.5\,\mathrm{R_e}$.\\
 & MW\_AGE\_GRADIENT & & dex/$\mathrm{R_e}$	& Mass-weighted age gradient of linear fit obtained within $1.5\,\mathrm{R_e}$.\\
 & MW\_AGE\_GRADIENT\_ERROR & & dex/$\mathrm{R_e}$	& Error on mass-weighted age gradient within $1.5\,\mathrm{R_e}$.\\
 & MW\_AGE\_ZEROPOINT & & & Mass-weighted age zeropoint of linear fit obtained within $1.5\,\mathrm{R_e}$.\\
 & MW\_AGE\_ZEROPOINT\_ERROR & & & Error on mass-weighted age zeropoint obtained within $1.5\,\mathrm{R_e}$.\\
 & LW\_Z\_GRADIENT & & dex/$\mathrm{R_e}$	& Light-weighted metallicity [Z/H] gradient of linear fit obtained within $1.5\,\mathrm{R_e}$.\\
 & LW\_Z\_GRADIENT\_ERROR & & dex/$\mathrm{R_e}$ & Error on light-weighted metallicity [Z/H] gradient within $1.5\,\mathrm{R_e}$.\\
 & LW\_Z\_ZEROPOINT & & & Light-weighted metallicity [Z/H] zeropoint of linear fit obtained within $1.5\,\mathrm{R_e}$.\\
 & LW\_Z\_ZEROPOINT\_ERROR & & & Error on light-weighted metallicity [Z/H] zeropoint obtained within $1.5\,\mathrm{R_e}$.\\
 & MW\_Z\_GRADIENT & & dex/$\mathrm{R_e}$	& Mass-weighted metallicity [Z/H] gradient of linear fit obtained within $1.5\,\mathrm{R_e}$.\\
 & MW\_Z\_GRADIENT\_ERROR & & dex/$\mathrm{R_e}$	& Error on mass-weighted metallicity [Z/H] gradient within $1.5\,\mathrm{R_e}$.\\
 & MW\_Z\_ZEROPOINT & & & Mass-weighted metallicity [Z/H] zeropoint of linear fit obtained within $1.5\,\mathrm{R_e}$.\\
 & MW\_Z\_ZEROPOINT\_ERROR & & & Error on mass-weighted metallicity [Z/H] zeropoint obtained within $1.5\,\mathrm{R_e}$.\\
\hline
\end{tabular}
\end{table*}

\begin{table*}
\contcaption{MaNGA {\sc firefly} VAC - Content of Catalogue}
\label{tbl:vac_content_cont}
\centering

\begin{tabular}{l p{0.1\linewidth} l p{0.16\linewidth} >{\hangindent=2em}p{0.4\linewidth}}
\hline\hline
HDU & Name & Dimensions & Units & Description\\
(1) & (2) & (3) & (4) & (5)\\
\hline
4 & SPATIAL INFORMATION (VORONOI CELL) & (5, 2800, 10735) & N/A, arcsec, arcsec, $\rm R_e$, deg & Spatial information, such as bin number, x-position, y-position and, in elliptical polar coordinates, radius (in units of effective radius) and azimuth for each Voronoi cell.\\
5 & SPATIAL INFORMATION (SPAXEL) & (80, 80, 10735) & N/A & 2D map of bin number.\\
6 & LIGHT-WEIGHTED AGE & (2, 2800, 10735) & $\rm \log_{10}\,(Gyr)$ & Light-weighted age, and associated error, derived from full spectral fit for each Voronoi cell.\\
7 & MASS-WEIGHTED AGE & (2, 2800, 10735) & $\rm \log_{10}\,(Gyr)$ & Mass-weighted age, and associated error, derived from full spectral fit for each Voronoi cell.\\
8 & LIGHT-WEIGHTED METALLICITY & (2, 2800, 10735) & N/A & Light-weighted metallicity [Z/H], and associated error, derived from full spectral fit for each Voronoi cell.\\
9 & MASS-WEIGHTED METALLICITY & (2, 2800, 10735) & N/A & Mass-weighted metallicity [Z/H], and associated error, derived from full spectral fit for each Voronoi cell.\\
10 & E(B-V) & (2800, 10735) & mag & E(B-V) derived from using High-Pass Filter (HPF) method for each Voronoi cell (See Goddard et al. 2017).\\
11 & STELLAR MASS & (4, 2800, 10735) & 0-1: $\rm \log_{10}\,(M_{\odot}/spaxel)$ 2-3: $\rm \log_{10}\,(M_{\odot})$ & Stellar mass, and associated error, derived from the full spectral fit for each Voronoi cell. Different to the global stellar mass. The first two channels give the stellar mass and error per spaxel, the last two channels give the total stellar mass and error of the Voronoi cell.\\
12 & STELLAR MASS REMNANT & (8, 2800, 10735) &  0-3: $\rm \log_{10}\,(M_{\odot}/spaxel)$ 4-7: $\rm \log_{10}\,(M_{\odot})$ & Mass of living stars, stellar remnants including white dwarf, neutron star, and black hole. The first four channels give the masses per spaxel, the last four channels give the total masses per Voronoi cell.\\
13 & SURFACE MASS DENSITY & (2, 2800, 10735) & $\rm \log_{10}\,(M_{\odot}\,kpc^{-2})$ & Surface Mass Density, and associated error, derived from the full spectral fit for each Voronoi cell.\\
14 & STAR FORMATION HISTORY & (3, 8, 2800, 10735) & $\rm \log_{10}\,(Gyr)$, N/A, N/A & Star formation history in each Voronoi cell reflected by the mass weight of each SSP that contributes to the observed spectra. The first channel of the array includes the age, metallicity [Z/H], and mass weights of each of the eight SSPs given in the second channel.\\
15 & SIGNAL-TO-NOISE & (2800, 10735) & S/N & Signal-to-noise calculated for each Voronoi cell in the wavelength range 4500-6000\AA.\\
\hline
\end{tabular}

\end{table*}

\section{Details from {\sc firefly} configuration tests}
\label{sect:config_plots}

\begin{figure*}
				\centering
				\includegraphics[width=18cm]{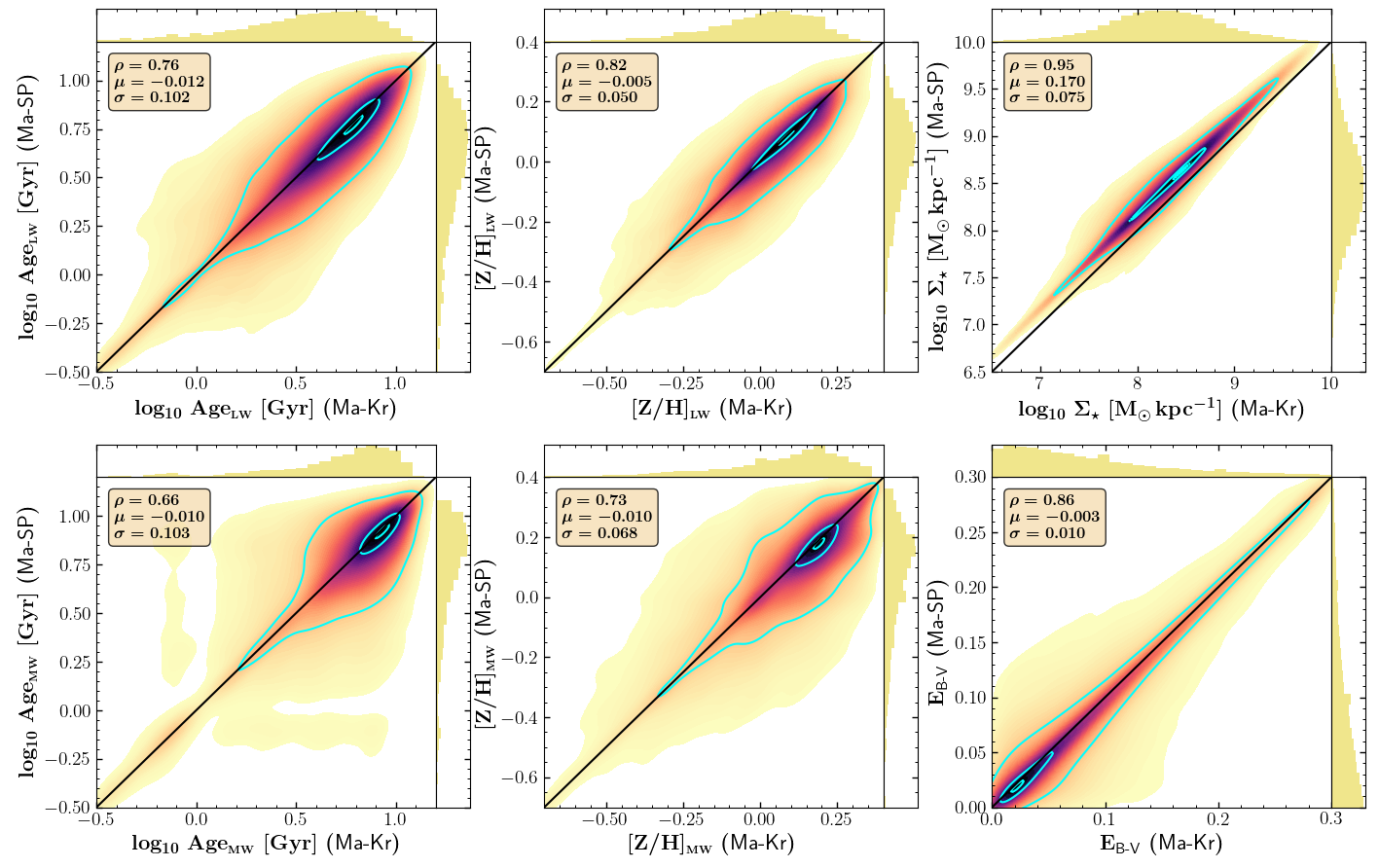}
    			\caption{Same as Fig. \ref{fig:mi_vs_ma} but comparing fits using MaStar SSPs with a Kroupa versus Salpeter IMF.}
    			\label{fig:imf}
			\end{figure*}

\begin{figure*}
				\centering
				\includegraphics[width=17cm]{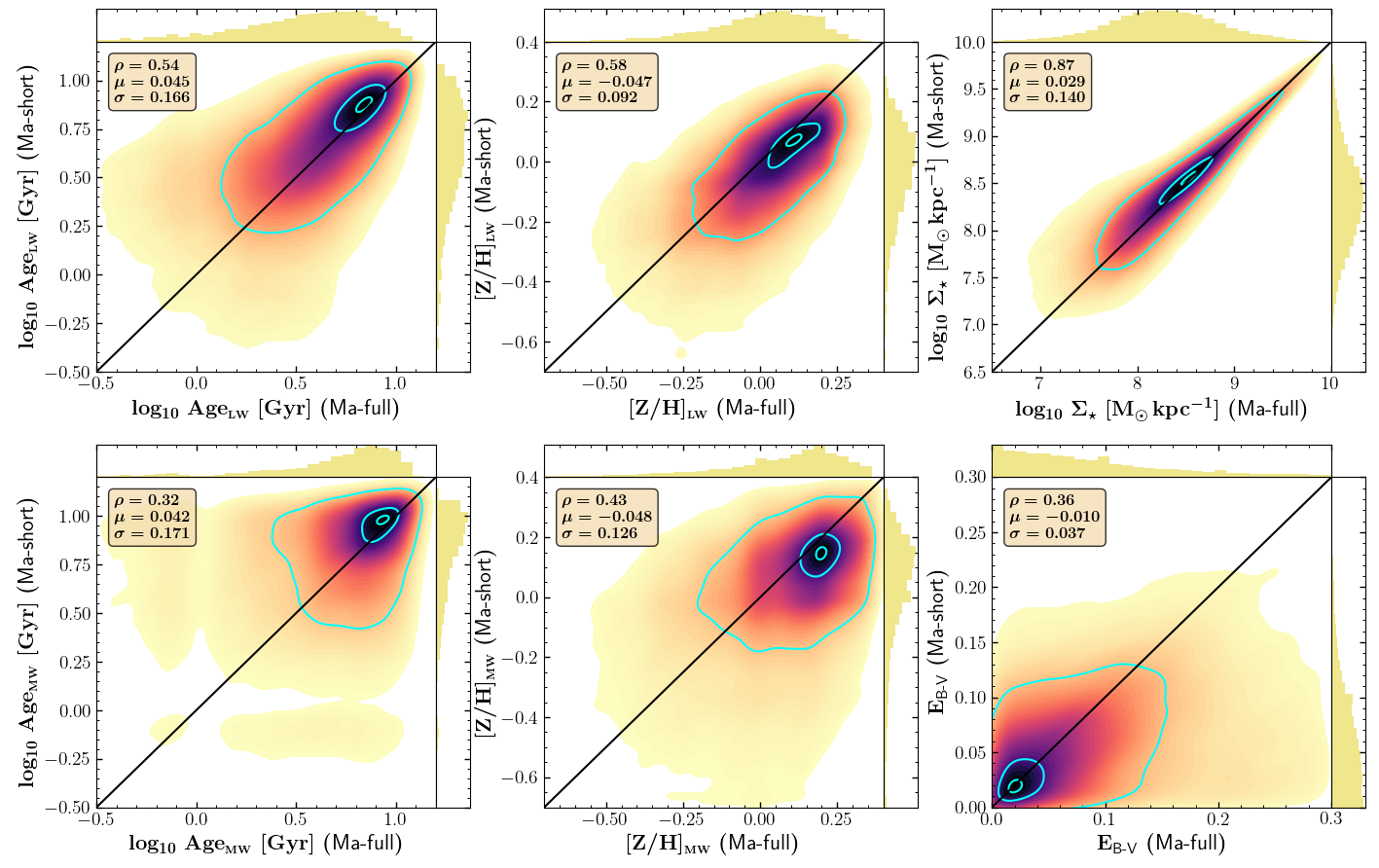}
    			\caption{Same as Fig. \ref{fig:mi_vs_ma} but comparing fits using Mastar with a full wavelength range versus MaStar with a short wavelength range.}
    			\label{fig:waverange_mi}
			\end{figure*}

\begin{figure*}
				\centering
				\includegraphics[width=17cm]{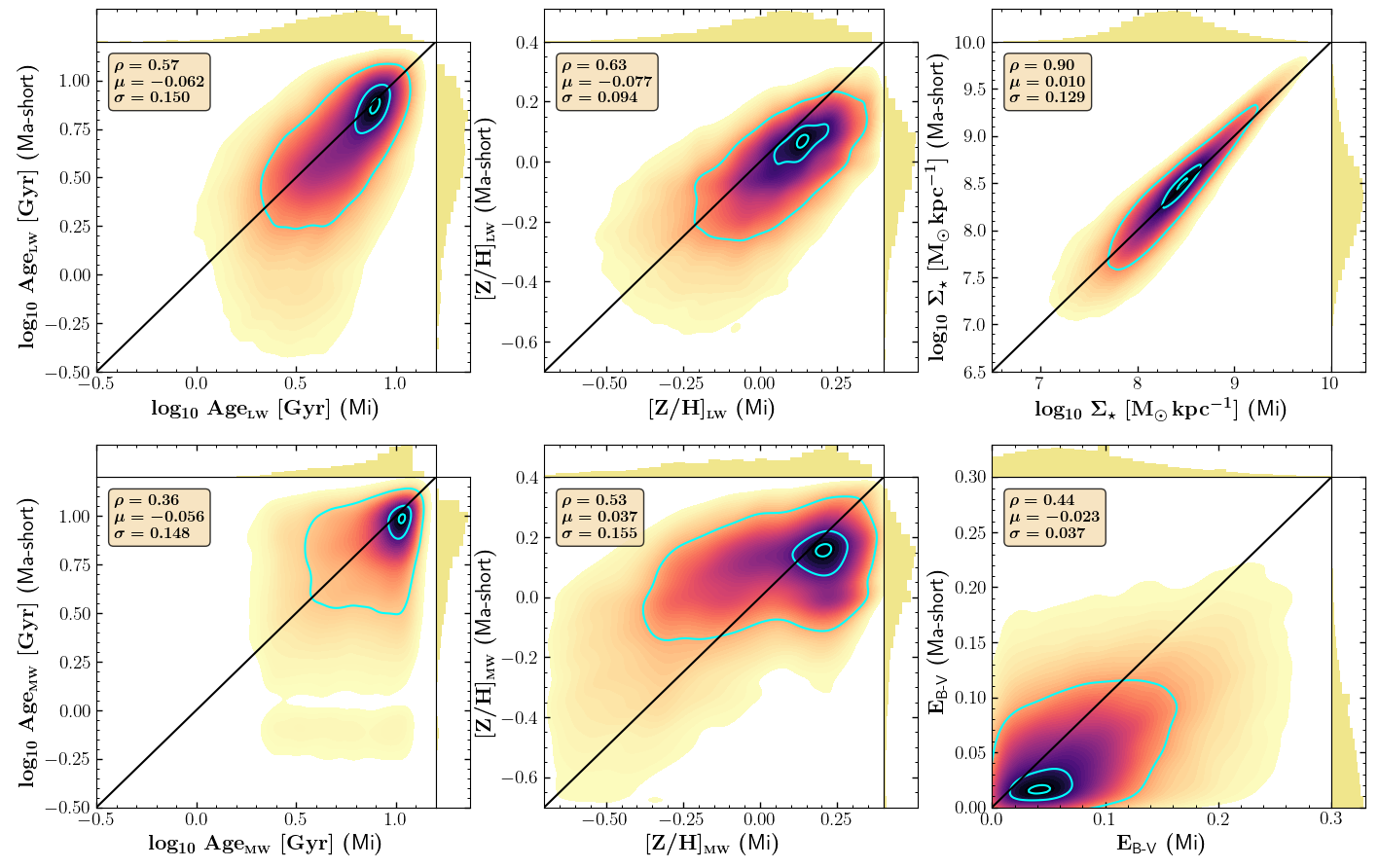}
    			\caption{Same as Fig. \ref{fig:mi_vs_ma} but comparing fits using M11-MILES and MaStar with the same short wavelength range.}
    			\label{fig:waverange_ma}
			\end{figure*}

\begin{figure*}
				\centering
				\includegraphics[width=17cm]{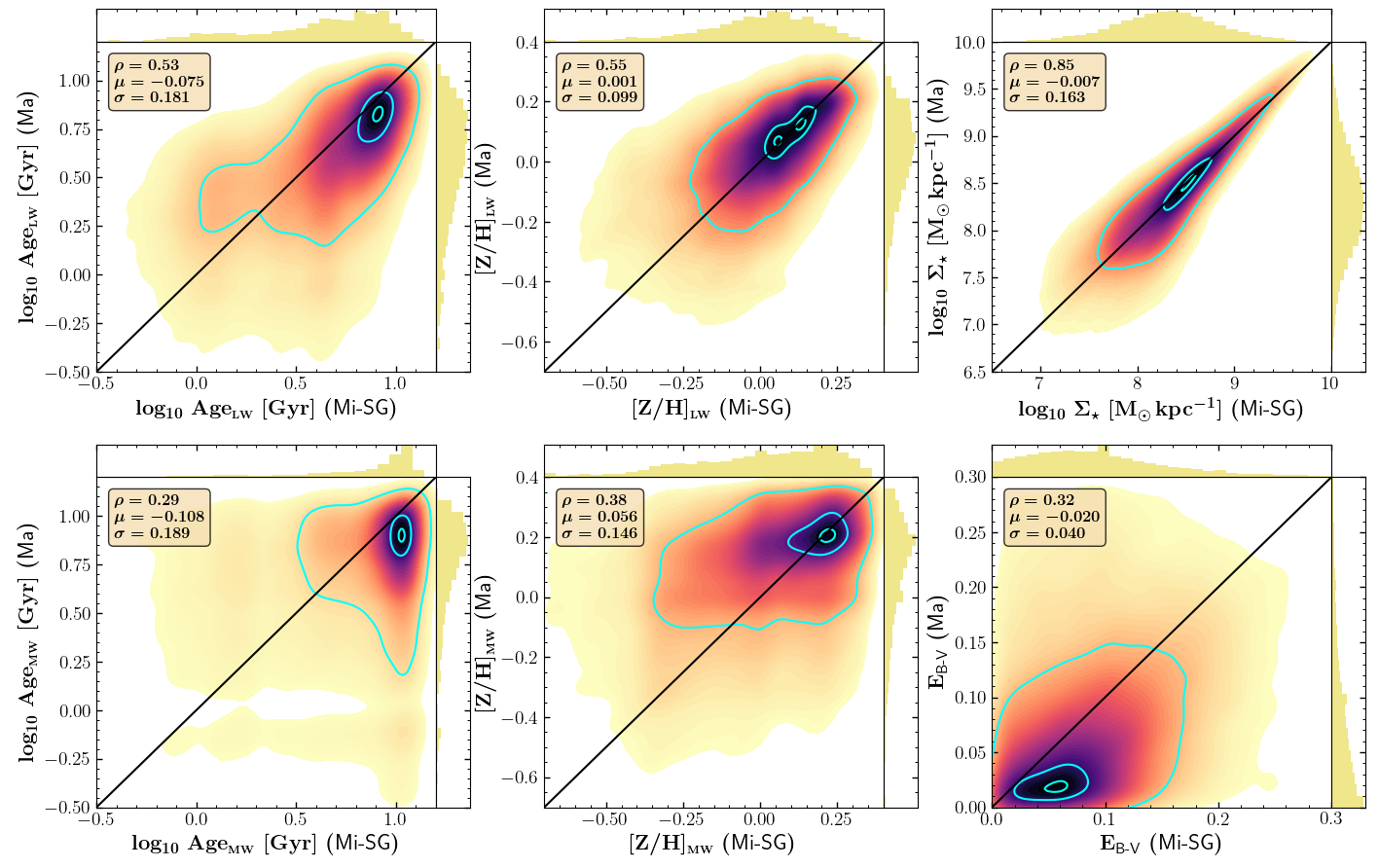}
    			\caption{Same as Fig. \ref{fig:mi_vs_ma} but comparing fits using M11-MILES squared grid models versus MaStar models.}
    			\label{fig:sg}
			\end{figure*}

\begin{figure*}
				\centering
				\includegraphics[width=17cm]{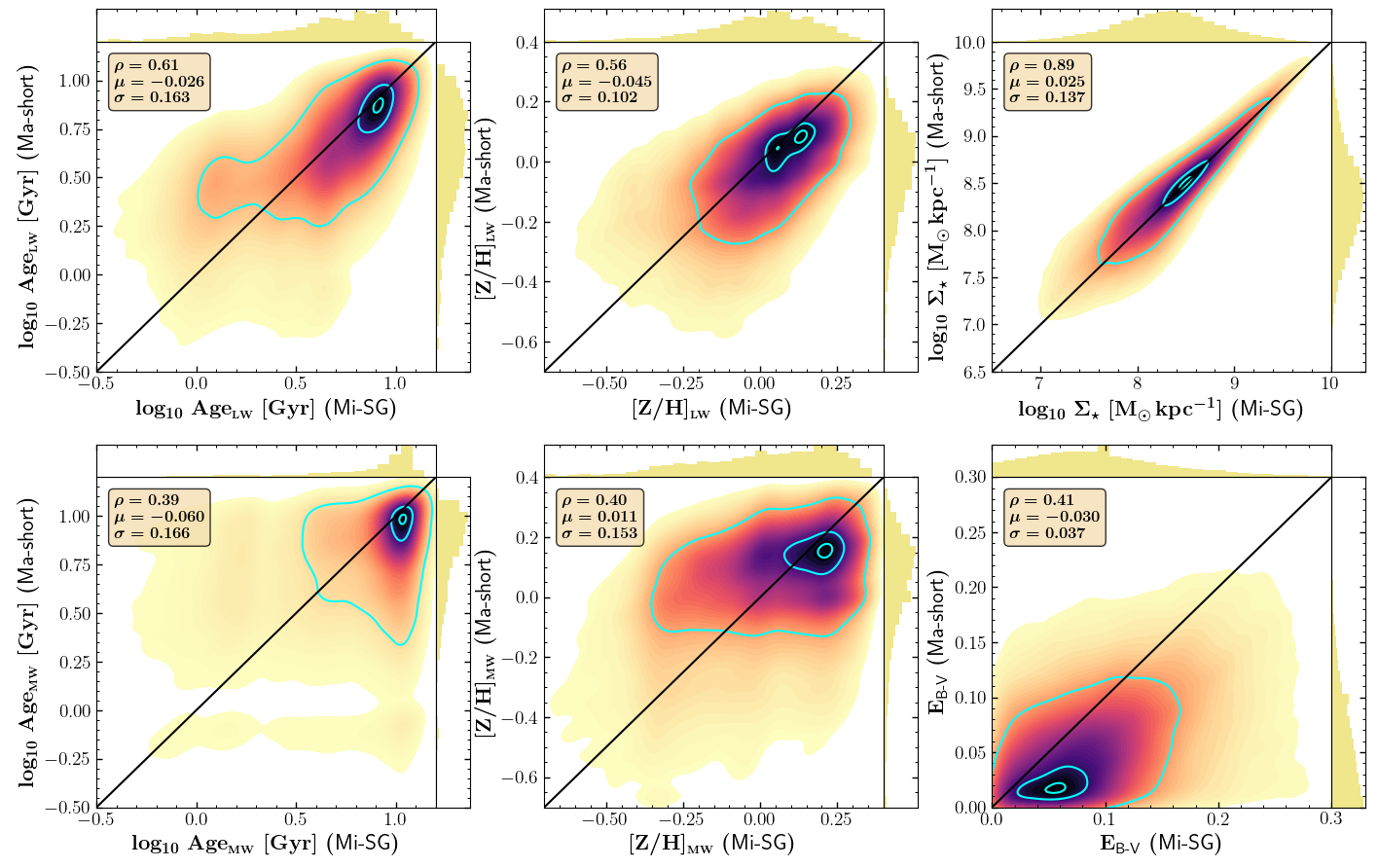}
    			\caption{Same as Fig. \ref{fig:mi_vs_ma} but comparing fits using M11-MILES squared grid models versus MaStar models with short wavelength range.}
    			\label{fig:sg_short}
			\end{figure*}

\begin{figure*}
				\centering
				\includegraphics[width=17cm]{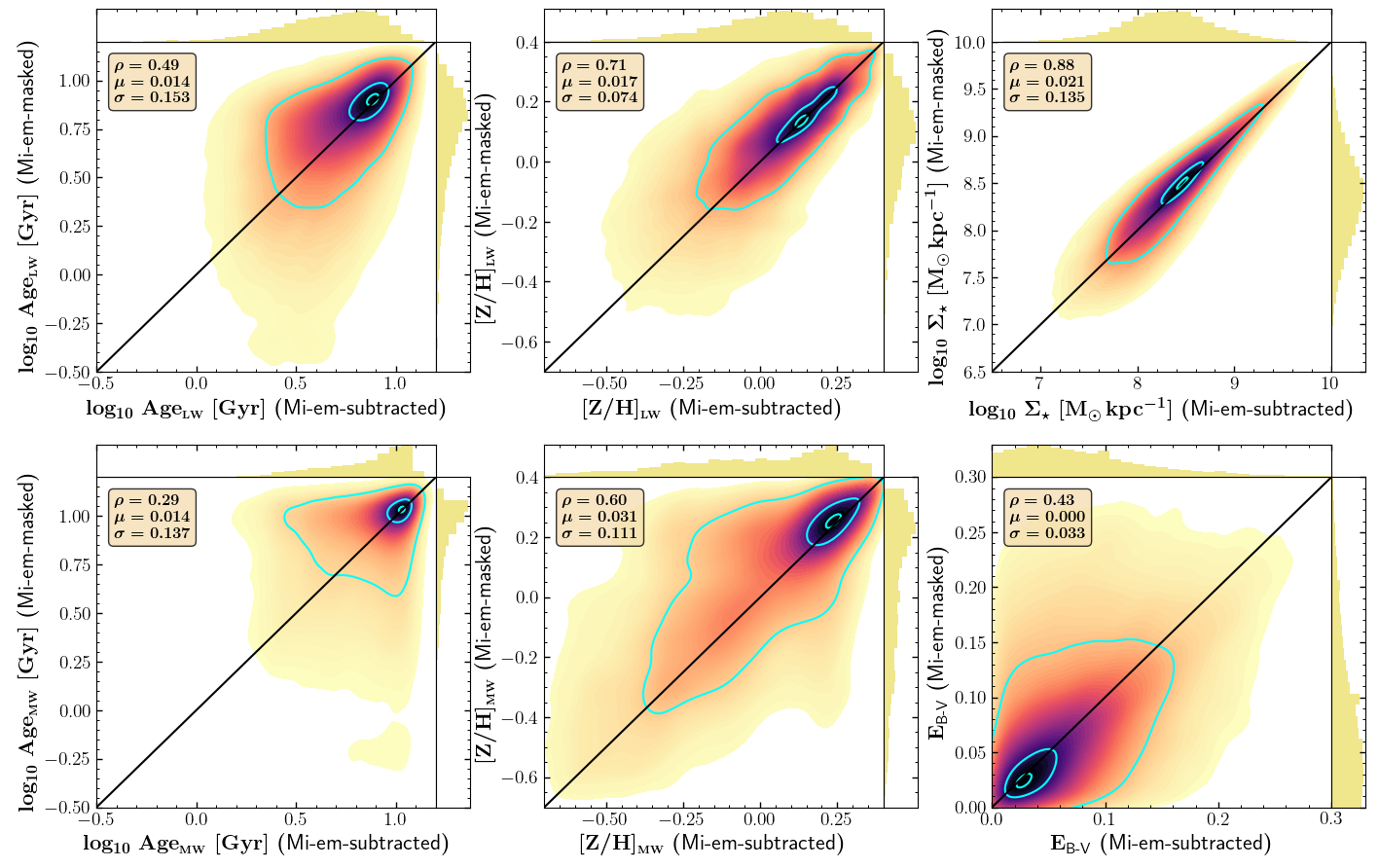}
    			\caption{Same as Fig. \ref{fig:mi_vs_ma} but comparing fits using M11-MILES models with emission lines subtracted versus emission lines masked.}
    			\label{fig:emlines_mi}
			\end{figure*}

\begin{figure*}
				\centering
				\includegraphics[width=17cm]{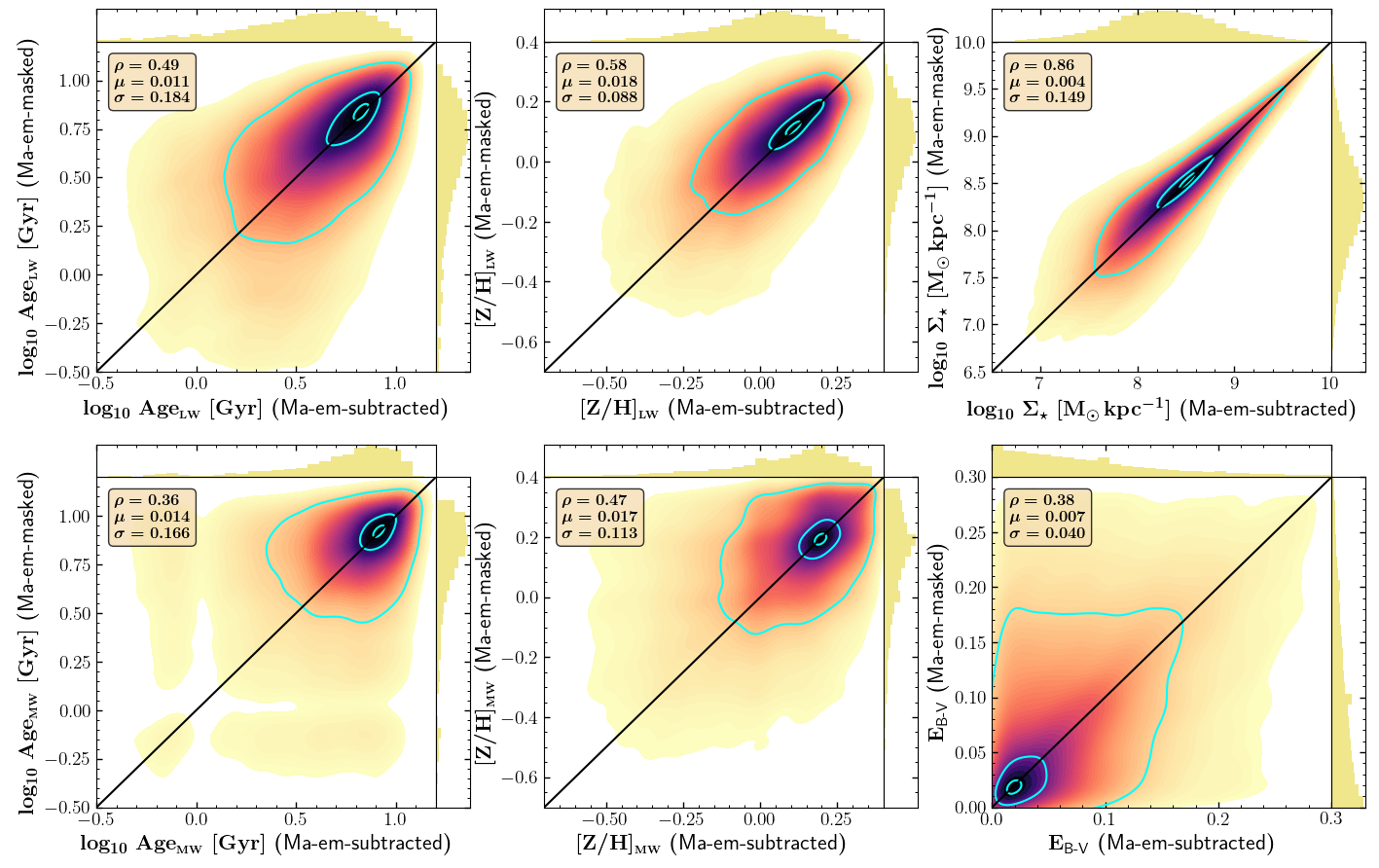}
    			\caption{Same as Fig. \ref{fig:mi_vs_ma} but comparing fits using Mastar models with emission lines subtracted versus emission lines masked.}
    			\label{fig:emlines_ma}
			\end{figure*}

We provide detailed plots from the configurations tests described in Sect. \ref{sect:config_tests} in Figs. \ref{fig:imf} to \ref{fig:emlines_ma}.

...

%%%%%%%%%%%%%%%%%%%%%%%%%%%%%%%%%%%%%%%%%%%%%%%%%%

% Don't change these lines
\bsp	% typesetting comment
\label{lastpage}
\end{document}